\documentclass{article}
\usepackage{tikz}
\usepackage{pgfplots}
\pgfplotsset{
  set layers,
  mark layer=axis tick labels,
  every axis/.append style={scale=1.2}
}
\usepackage[T1]{fontenc}
\usepackage[utf8]{inputenc}
\usepackage{color}
\usepackage{array}
\usepackage{multirow}
\usepackage{amsmath}
\usepackage{graphicx}
\usepackage{microtype}
\usepackage[unicode=true,
 bookmarks=false,
 breaklinks=false,pdfborder={0 0 1},backref=section,colorlinks=false]
 {hyperref}

\makeatletter
\providecommand{\tabularnewline}{\\}

\usepackage{arxiv}

\usepackage{url}
\usepackage{amsfonts}
\usepackage{nicefrac}
\usepackage{lipsum}
\usepackage{textcomp,mathcomp}
\usepackage{fullpage}
\usepackage{booktabs}
\usepackage{tabularx}
\usepackage{cite}
\usepackage{pdfpages}

\title{Computational Study of Non-isothermal Slag Eye Formation and its Effects on Ladle Refining}
\author{
  Anshuman Sinha \\
  Georgia Institute of Technology\\
  Atlanta, GA 30318 \\
  \texttt{anshs@gatech.edu} \\
   \And
 Amarendra K. Singh \\
  Indian Institute of Technology\\
  Kanpur, UP 10058 \\
  \texttt{amarendra@iitk.ac.in} 
}

\makeatother

\begin{document}
\maketitle 
\begin{abstract}
Ladle refining is one of the most important aspects of high-quality steel production. Ladle argon purging which facilitates the refining process also leads to the unwarranted opening of the slag cover known as Slag Eye-opening and has a deleterious effect on the quality of steel. Slag eye-opening has been analysed in past under isothermal conditions whereas ladle refining is a transient and non-isothermal operation. The current study deals with the modelling of slag-eye opening and its effects on ladle refining under non-isothermal conditions. The bubble plume is modelled with the help of Discrete Phase modelling (DPM) coupled with a discrete random walk model for including the particle level turbulence. Temperature-dependent thermophysical properties of slag are obtained from FactSage. Opening of slag-metal interface cools the slag-eye region, which causes changes in the thermophysical properties of the slag phase. These changes are then reflected in the flow characteristics of this complex fluid. The slag’s flow profile and eye formation are compared and explained between cold modelling techniques and actual ladle metallurgy. The consequences of changing thermophysical prop during ladle refining manifest in their influence on the overall mass transfer coefficient and the kinetics of desulfurization. This can be achieved without the requirement to solve computationally demanding species transport equations, thereby enhancing the practical efficiency of this approach.
\end{abstract}

\section{Introduction}
\label{sec:intro}
\begin{figure}[h!]
\centering 
\includegraphics[width=1\linewidth]{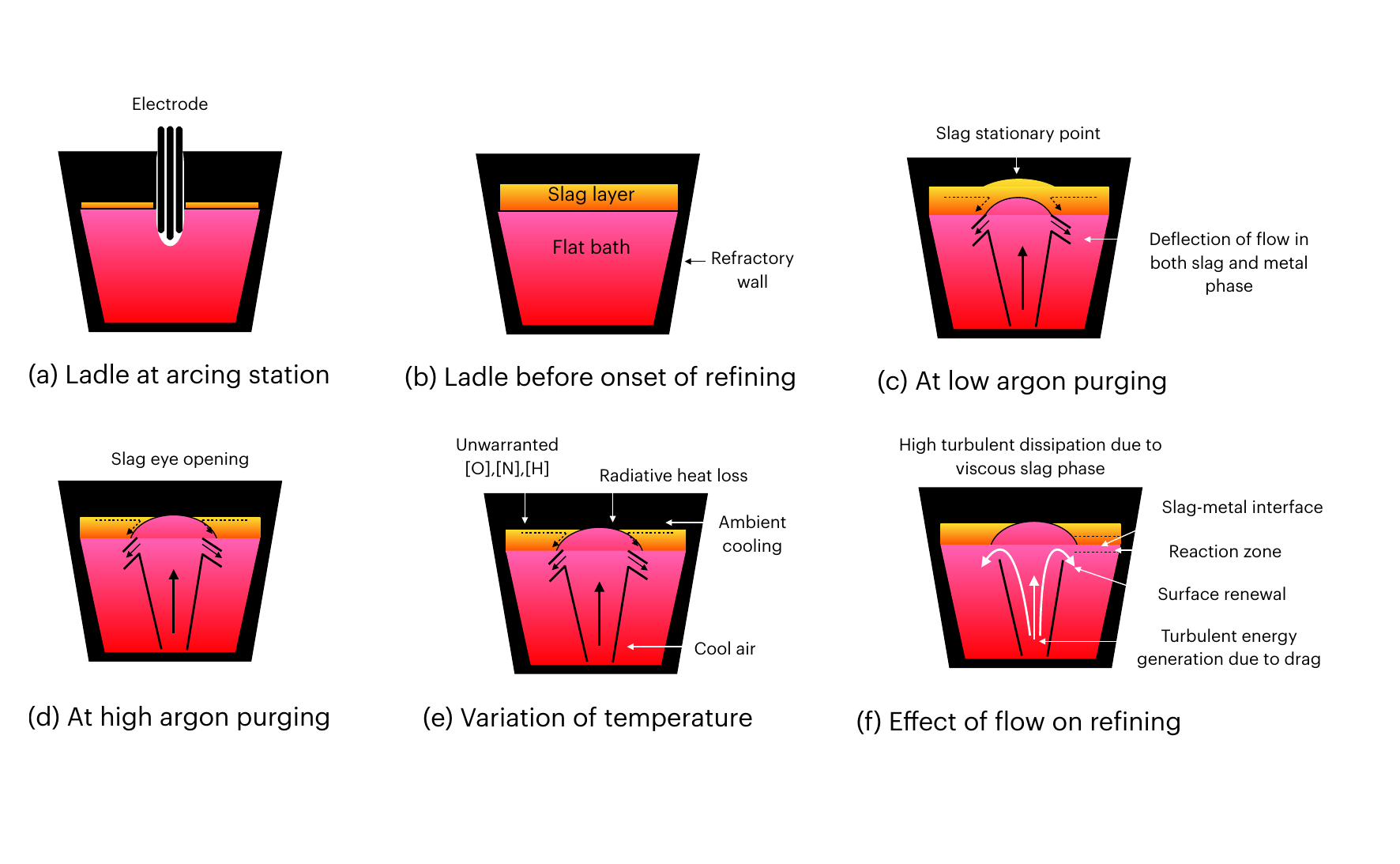}
\caption{Schematic of step-by-step process flow taking place in the ladle furnace
from flat bath condition after EAF/BOF operations to highly turbulent
refining condition at the ladle furnace operation unit. }
\label{fig:steps-of-laddle-furn} 
\end{figure}

Crude steel requires further refining and alloying to obtain steel
of desired composition and cleanliness and refining of steel through
one or more secondary steelmaking such as ladle refining, RH degassing
etc. In the ladle refining process, inert gas, usually argon, is purged
into the molten metal bath from the bottom through tuyeres or porous
plugs. One of the important refining operations is the ladle furnace
(LF) refining operation, in which argon gas stirring is essential
in determining and improving the quality of liquid steel \cite{Zhang2003}.
However, the strong purging conditions required for ladle refining,
also lead to some unwarranted disadvantages in terms of opening the
slag cover, which is also termed as slag-eye opening. Fig.~\ref{fig:steps-of-laddle-furn},
shows schematics of various ladle operations in a step-by-step manner
which leads to the desired final steel composition. Here, we can see
the various purging conditions giving rise to the bubble-driven liquid
metal plume, which due to its buoyancy-driven momentum, provides the
necessary circulations needed for the renewal of the slag metal interface
and temperature and composition homogenization of the liquid metal
bath. As shown in Fig.~\ref{fig:steps-of-laddle-furn}(e), the slag
eye opens the way for the pickup of unwarranted elements like Oxygen,
Hydrogen and Nitrogen as well as exposes the hot metal to the ambient
for heat loss \cite{Zhang2003,Ghosh2000}. Hence, minimising slag
eye-opening is essential, and a proper understanding of its effects
on the overall refining process is important. The understanding of
flow fields and changing thermophysical properties pertaining to the
interfacial region (as shown in Fig.~\ref{fig:steps-of-laddle-furn}(f),
due to the corresponding heat loss during the refining process helps
in quantifying the slag eye size and also provides an insight towards
the efficiency of the interfacial refining reactions. 

Over the past few decades, many experimental studies and measurements
have tried to study the flow field and slag eye formation in view
of dynamic similarity in the gas-stirred ladle \cite{Xie1992,Yonezawa1999,Mazumdar2004,Krishnakumar2008,Wu2010,Krishnapisharody2006,Liu2017}. These experiments were done on cold water models with an overlying
oil phase depicting the slag phase. These experiments were conducted
with variable bath heights and air flow rates in order to study their
effects on slag eye radius and flow properties. The experiments provide
quality data for the validation of complex multi-physics mathematical
models. Most of the work \cite{Volkova2003,Camdali2006,Zimmer2008,Livshits2011,Glaser2011,Kabakov2013,Seshadri2016,FarreraBuenrostro2019}
related to non-isothermal studies in the ladle has been done in order
to quantify the heat losses during the ladle furnace operations. While
these models successfully predict the contribution of various factors
towards heat loss, they don’t present a study on the effects of these
on flow fields or refining conditions. Gonzalez et. al. \cite{Gonzalez2017}
studied the thermal and chemical homogenization using argon injection
sequences with a non-isothermal 4-phase 3D mathematical model. They
assumed the thermophysical properties to be held constant throughout
the transient simulation, due to which the effects of these parameters
could not be studied. Along with the extensive experimental studies, there have been, many
physical and numerical modelling studies \cite{Han2001,TafaghodiKhajavi2010,Liu2018,Guo2002,Thunman2007,Lv2017,Valentin2009,Liu2011}
conducted to study the fluid flow behaviour and open-eye formation
in the gas-stirred ladle. Over the past decade, the studies were mainly
focused on investigating the effect of gas flow rate on the open-eye
formation, and relatively few studies on the effect of slag properties
(density, dynamic viscosity and upper layer thickness) on the open-eye
size through both physical and numerical modelling, the topic which
we have taken up as a central theme to our present work.

Furthermore, few investigations in the past have included the effects
that the top slag layer has on flow properties \cite{Li2008,Li2015,Li2016,Li2017,Singh2016}.
Most of these works have reported the effects of these properties
on mixing time and have done parametric studies of argon flow rate,
bath height, nozzle position, on homogenization of the bath. All these
models do not incorporate the transient behaviours of slag phase properties
via incorporating equilibrium chemistry models and it’s the corresponding
effect on the composition of the bath. Ramasetti \cite{Ramasetti2019}
tried to explain the effects of slag properties on slag eye area with
the help of multiple experiments and CFD models with various top layer
fluids in a cold model setup, but did not model the transient behaviour
of the slag phase in non-isothermal conditions. While work from Jonson
(\cite{Jonsson1996,Jonsson1998}) implements the 2D transient model
for slag’s properties in a ladle but did not discuss any results pertaining
to the transient effects of slag properties on fluid flow, slag eye
or the refining process in the ladle.

The present work investigates the effects of the non-isothermal condition
of ladle refining which has not yet been addressed in the context
of ladle furnace refining and shows the importance of these transient
effects on refining after being validated by industrial experimental
results. The work focuses on modelling gas-agitated multiphase fluid
flow inside the ladle and describes the phenomenon of ladle slag eye
formation under non-isothermal conditions which affects the thermophysical
and flow properties of the slag phase and finally the refining of
molten steel. A 3-Dimensional Multiphysics Eulerian-DPM coupled model
is set up with varying thermophysical properties of the slag phase
according to the thermodynamic relations. The application of the current
model is presented in view of the effects of external argon purging
on fluid flow conditions and the overall ladle refining process which
can be used directly for industrial application or can serve as a
digital twin to make surrogate models for fast model predictions.
The work concludes that the non-isothermal modelling presents a more
accurate prediction of refining behaviours with increasing rates of
desulfurization at higher gas flow rates. Our results show the slowing
down of mass transfer phenomena with time as a result of the transient
thermochemical behaviour of the ladle during operation. The work opens
the door to better and more accurate parametric optimization of the
process variables in order to increase the productivity of the metallurgical
plants. The application of this is shown with respect to ladle desulfurization
by our model in the result section (Section~\ref{sec:Results-and-discussions}),
which can be opted for industrial use with appropriate scaling.

\section{Problem description}
\label{sec:Prob-discri}
The steel ladle in the current model as described in Table~\ref{tab:dim-ladle},
operates with a central porous purging unit for argon purging. The
use of a central porous plug gives the advantage while studying the
effects of flow properties on slag eye formation and ladle refining-related
phenomena. This is taken so, to ensure that none of the spatial-temporal
variations in the slag phase occurs due to off-centric purging. The
ladle is set for refining after the primary steelmaking operation.

\begin{figure}[h!]
\centering 
\includegraphics[width=1\linewidth]{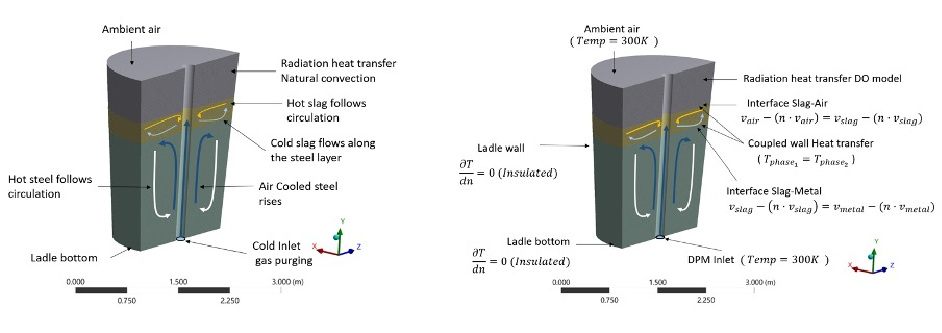} 
\caption{Shows the various boundary conditions and their implementation. (a)
Physical boundary conditions and (b) their mathematical interpretations.}
\label{fig:BCs} 
\end{figure}

\begin{table}
\centering
\caption{Dimensions of the ladle used in the current work.}
\begin{tabular}{ccccccc|}
\hline 
{Weight ($kg$)} & Density ($kg/m^{3}$) & Volume ($m^{3}$) & $\lambda_{1}$ & $\lambda_{2}$ & Height ($m$) & Diameter ($\mu$$m$)\tabularnewline
\hline 
120000 & 7890 & 15.23 & 1.8 & 1.16 & 2.378 & 2.64\tabularnewline
\hline 
\end{tabular}
\label{tab:dim-ladle}
\end{table}

The physical description of the problem is described in Fig.~\ref{fig:BCs}(a).
The various parts of the ladle furnace, along with its components
are shown in the figure. The outward tapered ladle consists of liquid
steel with overlying slag in the ambient environment. A central purging
nozzle is used in order to provide necessary flow circulation with
argon gas. The liquid steel and slag system is exposed to cold ambient
and argon gas which gradually cools the content of the ladle and thermal
gradients develop inside these phases. The thermophysical parameters and interfacial tension between the
continuous phases is described in Table~\ref{tab:phy-props-sim}
and Table~\ref{tab:surf-tension-phases}. The argon flow in the current
work is assumed to be in the form of discrete bubbles, the properties
of which are described in Table~\ref{tab:phy-prop-dpm} and the properties
of DPM Inlet as shown in Fig.~\ref{fig:BCs}(b) is described in Table~\ref{tab:dpm-particle-inlet-charct}.

\begin{table}
\centering
\caption{Physical parameters for the simulations \cite{Ramasetti2019}}
\begin{tabular}{ccc}
\hline 
Physical properties at $1873K$ & Value & Unit\tabularnewline
\hline 
Density of liquid steel  & $7880$ & $kg/m^{3}$\tabularnewline
Viscosity of liquid steel  & $0.0055$ & $Pa\,s$\tabularnewline
Density of slag & $3500$ & $kg/m^{3}$\tabularnewline
Viscosity of slag & $0.104$ & $Pa\,s$\tabularnewline
Thermal conductivity of slag phase \cite{Lejeune1995} & $0.115$ & $W/m/K$\tabularnewline
Thermal conductivity of steel phase \cite{Launder1972} & $320$ & W/m/K\tabularnewline
Density of Argon & $0.8739$ & $kg/m^{3}$\tabularnewline
Viscosity of Argon gas & $2.261\times10^{5}$ & $Pa\,s$\tabularnewline
\hline 
\end{tabular}
\label{tab:phy-props-sim}
\end{table}

\begin{table}
\centering
\caption{Surface tension modelling between continuous phases \cite{Ramasetti2019}}
\begin{tabular}{ccc}
\hline 
Physical properties at $1873K$ & Value & Unit\tabularnewline
\hline 
Steel-air & $1.82$ & $N/m$\tabularnewline
Steel-slag & $1.15$ & $N/m$\tabularnewline
Density of slag & $0.58$ & $N/m$\tabularnewline
\hline 
\end{tabular}
\label{tab:surf-tension-phases}
\end{table}

\begin{table}
\centering
\caption{Physical parameters of the discrete bubble phase}
\begin{tabular}{ccc}
\hline 
Physical properties at $1873K$ & Value & Unit\tabularnewline
\hline 
Bubble Size & $4$ & $mm$\tabularnewline
Bubble density & $1.68$ & $g/cm^{3}$\tabularnewline
Bubble viscosity & 0.001 & $Pa\,s$\tabularnewline
Surface tension & $0.015$ & $N/m$\tabularnewline
\hline 
\end{tabular}\label{tab:phy-prop-dpm}
\end{table}

\begin{table}
\centering
\caption{DPM particle inlet characteristics}
\begin{tabular}{ccc}
\hline 
Physical properties at $1873K$ & Value & Unit\tabularnewline
\hline 
Particle diameter distribution & Uniform & $-$\tabularnewline
Initial velocity & $0$ & $m/s$\tabularnewline
Number of holes & $5$ & $-$\tabularnewline
Injection type & Inlet surface & Normal to surface\tabularnewline
Mass flow rate & $4.65\times10^{-5}$ & $kg/s$\tabularnewline
\hline 
\end{tabular}
\label{tab:dpm-particle-inlet-charct}
\end{table}

The physical model is converted into a mathematical form in order
to solve the interest variables. The fundamental first step towards
this solution procedure is to discretize the domain of interest into
several small and strategically shaped domains called cells. Moreover,
the stencil which is used to discretize the domain into cells is called
mesh. In the above-described problem, meshing is to be given proper
attention as multiple physics are taking place in the domain, which
may have different and competing requirements of cell sizes for accurate
prediction. A full 3D model is constructed to understand the behaviour
and associated flow field and the corresponding formation of the slag-eye.
The random walk turbulence model for DPM modelling limits the applicability
of the axisymmetric condition in the modelling sense. Moreover, the
temperature loss caused by the slag eye is highly dependent on the
curvature of the exposed molten metal and slag surfaces due to the
inclusion of radiation heat transfer models. Assuming an axisymmetric
model would lose out on these details and would not give an accurate
picture of non-isothermal conditions.

The set of assumptions have been made in the present calculations,
which are as follows: 
\begin{enumerate}
\item The geometry of the ladle is assumed to be tapered cylindrical, as
mentioned in various previous literature. The dimensions are calculated
based on (the ‘$l=H/r$’ ratio, ladle tonnage ($W$), and taper angle
(or $l=r_{2}/r_{1}$ ratio)) as given in Table~\ref{fig:steps-of-laddle-furn}. 
\item At the start of the simulation, the liquid steel bath and the slag
phase in the ladle are assumed at a constant and homogenized temperature
of $1873K$ ($1600$ $^\circ{C}$). 
\item Liquid steel, slag, and argon have been considered to behave like
incompressible Newtonian fluids. Argon, being gas, is compressible.
However, in the context of ladle processing, researchers in the past
have treated argon as both compressible \cite{Geng2010,Llanos2010}
as well as incompressible \cite{Li2008,Zhang2012,Haojian2018,Batchelor1967,Roscoe1952}.
We have treated argon as an incompressible fluid as the flow conditions
in the ladle have a low Mach number, and the compressibility factor
for argon is acceptably small for the pressure stratification due
to the height of the liquid steel.
\item The ambient is assumed to be at $300\,K$. The heat transfer from
the ladle is only considered from the top of the ladle. The refractory
sidewalls are assumed to be adiabatic as the heat loss from the sidewalls
would not significantly affect the change in slag’s temperature ($T_{slag}$). 
\item The ladle refining process is articulated with the help of convective
mass transfer co-efficient only. The explicit interfacial chemical
reaction between steel and slag phase is not considered separately. 
\item A full 3D model is constructed to understand the behaviour and associated
flow field and the corresponding formation of the slag-eye. The random
walk turbulence model for DPM modelling limits the applicability of
the axisymmetric condition.
\item The composition of all the phases is assumed to be constant to limit
the computation cost of this 3D transient model.
\end{enumerate}

\section{Numerical modelling}
\label{sec:num-modeling}
The physical model as described in Fig.~\ref{fig:BCs}(b) is implemented
in a three-dimensional transient mathematical model for argon gas
stirred ladle which accounts for the steel, slag, and argon phases
has been developed. A set of Navier--Stokes equations with the incorporation
of the volume-of-fluid (VOF) function has been solved to investigate
the dynamic behaviour of the three phases. The boundary conditions
of the problem statement are shown in Fig.~\ref{fig:BCs}(b). The
following sets of transport equations \cite{Marsh1981}are solved. 

\subsection{Governing equations}

\paragraph*{Continuity equation}

\begin{equation}
\frac{\partial\rho}{\partial t}+\nabla\cdot(\rho\boldsymbol{v})=0
\label{eq:conti}
\end{equation}

\paragraph*{Momentum equation}

\begin{equation}
\frac{\partial\rho}{\partial t}+\nabla\cdot(\rho\boldsymbol{vv})=-\nabla p+\nabla\cdot\left[\mu_{e}\left(\nabla\boldsymbol{v}+\nabla\boldsymbol{v}^{T}\right)\right]+\rho\boldsymbol{g}
\label{eq:mom}
\end{equation}

where $p$ is the local pressure, $\boldsymbol{g}$ is the acceleration
due to gravity, $\boldsymbol{v}$is the local velocity and $\mu_{e}$
is the effective viscosity. The effective viscosity is calculated
as the sum of dynamic and turbulent viscosities.

\paragraph*{Energy equation}

\begin{equation}
\frac{\partial\rho}{\partial t}+\nabla\cdot(\boldsymbol{v}\left[\rho E+p\right])=\nabla\cdot\left[k_{eff}\nabla T-\sum_{j}h_{j}\boldsymbol{J}_{j}+\boldsymbol{\tau}\cdot\boldsymbol{v}\right]+S_{E}
\label{eq:engy}
\end{equation}

$k_{eff}$ is effective conductivity of the medium in which energy
is to be propagated, $h$ is enthalpy or the heat contednt of the
medium, $S_{E}$ is the source term which can be used to define external
sources like heating elemnets. The term $E=h+\frac{p}{\rho}+\frac{\left|v\right|^{2}}{2}+k_{eff}$
is defined as the net energy of the system as defined in equation
Eq.~\ref{eq:engy}

\paragraph{DPM equation}

\begin{equation}
\frac{d\rho}{dt}=F_{D}\left(\boldsymbol{u}-\boldsymbol{u}_{p}\right)+\frac{\boldsymbol{g}\left(\rho_{p}-\rho\right)}{\rho_{p}}+\boldsymbol{F}
\label{eq:dpm}
\end{equation}

where $\boldsymbol{F}$ is an additional acceleration (force/unit
particle mass) term, $F_{D}\left(\boldsymbol{u}-\boldsymbol{u}_{p}\right)$
is the drag force per unit particle mass and $F_{D}=\left(18\mu/24\rho_{p}d_{p}^{2}\right)\left(C_{D}Re/24\right)$,
and corresponding $\mathrm{Re}=\rho d_{p}\left|\boldsymbol{u}-\boldsymbol{u}_{p}\right|/\mu$
is the Reynolds number with respect to the continuous phase. Here,
$\boldsymbol{u}$ is the fluid phase velocity, $\boldsymbol{u}_{p}$
is the particle velocity, $\rho$ is the fluid density, $\rho_{p}$
is the density of the particle, and $d_{p}$ is the particle diameter.
$\mathrm{Re}$ is the relative Reynolds number. 

\paragraph*{Radiation transfer equation}

The contribution to heat transfer from radiation was calculated by
using the non-grey discrete ordinates (DO) radiation model. In this
model, we compute the heat transfer in the direction of heat transfer
rays $\boldsymbol{s}$, along this direction the radiative transfer
equation for spectral intensity $I_{\lambda}(\boldsymbol{r},\boldsymbol{s})\boldsymbol{s}$
can be written as follows: 

\begin{equation}
\nabla\cdot I_{\lambda}(\boldsymbol{r},\boldsymbol{s})\boldsymbol{s}=\left(a_{\lambda}+\sigma_{\lambda}\right)I_{\lambda}(\boldsymbol{r},\boldsymbol{s})+a_{\lambda}n^{2}I_{b\lambda}+\frac{\sigma_{s}}{4\pi}\int I_{\lambda}(\boldsymbol{r},\boldsymbol{s}^{'})\,\phi(\boldsymbol{r},\boldsymbol{s}^{'})\,d\Omega^{'}
\label{eq:rad-tranf}
\end{equation}

The four terms in Eq.~\ref{eq:rad-tranf} represent the rate of increase
in radiation intensity, loss by absorption and out-scattering, gain
by emission, and gain by in-scattering, respectively. The present
study applies the grey-band model for eight discrete bands {[}45{]}
and uses the averaged absorption coefficient for each wavelength band.
The absorption coefficient al in Eq.~\ref{eq:rad-tranf} was estimated
from the measured transmittance and reflectance \cite{Marsh1981,Lejeune1995}. 

\paragraph*{Turbulence $\kappa-\epsilon$ model equations}

\begin{equation}
\frac{\partial}{\partial t}\left(\rho\kappa\right) + \nabla\cdot\left(\rho\kappa\boldsymbol{u}_{i}\right) = 
\nabla\cdot(\left[\mu+\frac{\mu_{t}}{\sigma_{k}}\nabla(\epsilon)\right])
+G_{k}+G_{b}-\rho\boldsymbol{\epsilon}-Y_{m}+S_{k}
\label{eq:k}
\end{equation}

\paragraph*{Dissipation equation}

\begin{equation}
\frac{\partial}{\partial t}\left(\rho\epsilon\right) + \nabla\cdot\left(\rho\epsilon\boldsymbol{u}_{i}\right) = 
\nabla\cdot(\left[\mu+\frac{\mu_{t}}{\sigma_{k}}\nabla(\epsilon)\right])
+C_{1\epsilon}\frac{\epsilon}{\kappa}\left(G_{\kappa}+C_{3\epsilon}G_{b}\right)-C_{2}\epsilon\rho\frac{\epsilon^{2}}{\kappa}+S_{e}
\label{eq:eps}
\end{equation}

Where the terms $G_{k}$and $G_{b}$ represent the generation of kinetic
energy due to turbulence from the mean velocity gradient and the generation
of kinetic energy due to turbulence from the effects of buoyancy,
respectively. Similarly, $S_{k}$ and $S_{\epsilon}$ are sink terms
in the respective equations which counter the generation of energy
or dissipation by eliminating the energy to other smaller scales of
motion. $\mu_{t}=\rho C_{\mu}\left(\kappa^{2}/\epsilon\right).$ The
turbulent viscosity is calculated as $\mu_{t}$ by using the ‘$\kappa$’
and ‘$\epsilon$’ from Eq.~\ref{eq:k} and Eq.~\ref{eq:eps}
respectively. In these equations many constants are used whose values
are as follows: $C_{1}\epsilon=1.44$, $C_{2}\epsilon=1.92$, $\sigma_{k}=1.0$,
$C_{3}\epsilon=1$.0, and $\sigma_{\epsilon}=1.3$ \cite{Launder1972}.

\paragraph*{Volume-of-Fluid (VOF) fomulation}

The VOF formulation assumes that the various phases present in the
system are not interpenetrating. While the interaction between the
phases is based on the modelling of the interfaces between the phases.
Therefore, each phase in a cell is represented by its volume fraction.
Volume fractions are derived by solving the continuity equation for
each phase given in equations below.

\begin{equation}
\frac{\partial\alpha_{q}}{\partial t}+\left(\boldsymbol{v}\cdot\nabla\right)\frac{\partial\alpha_{q}}{\partial t}=0
\label{eq:vof}
\end{equation}

\begin{equation}
\alpha_{g}+\alpha_{l}+\alpha_{s}=0\label{eq:vof-sum-frac}
\end{equation}
where $\alpha$ is the volume fraction value of $g$, $l,$ and $s$
for argon gas, steel, and slag phases.

\begin{equation}
\rho=\alpha_{g}\rho_{g}+\alpha_{l}\rho_{l}+\alpha_{s}\rho_{s}
\label{eq:vof-den}
\end{equation}
The density of the mixture is calculated by Eq.~\ref{eq:vof-den}
and used in Eq.~\ref{eq:conti}--\ref{eq:dpm} and Eq.~\ref{eq:k}--\ref{eq:eps}.

\paragraph*{Equilibrium model for thermophysical properties}

The most important factors governing the rheological properties of
slag are its density $\rho_{slag}$, viscosity of the liquid phase
$\eta_{eff}$ and the fraction of crystals $\phi$. For the equilibrium
model, we have considered the following slag system is shown in Table~\ref{tab:props-slag}
for our study. The equilibrium phase diagrams along with the apparent
viscosity both are calculated by the ‘Equllib’ and ‘Viscosity’ packages
respectively of FactSage software \cite{bale2002factsage}. 

Later the data from these two packages are used to calculate the density
and viscosity of the slag phase at various slag operating temperatures,
in order to get the relations between temperature and effective viscosity
from Eq.~\ref{eq:eff-visco} and temperature and slag density from
Eq.~\ref{eq:rho-slag}.

\begin{table}
\centering
\caption{Composition of the slag used in our current work \cite{Sinha2020}}
\begin{tabular}{ccccccc|}
\hline 
\multirow{1}{*}{$Al_{2}O_{3}$($\left[wt\%\right]$)} & $SiO_{2}$($\left[wt\%\right]$) & $CaO$($\left[wt\%\right]$) & $MgO$ ($\left[wt\%\right]$) & Temp. ($K$) & Viscosity ($Pa\,s$) & density ($kg/m^{3}$)\tabularnewline
\hline 
120000 & 7890 & 15.23 & 1.8 & 1.16 & 2.378 & 2.64\tabularnewline
\hline 
\end{tabular}
\label{tab:props-slag}
\end{table}

The slag’s thermophysical properties are strongly dependent on the
phase composition of the slag phase at the different operating temperatures. FactSage software \cite{bale2002factsage} is used to obtain the solid fractions of the above slag composition at different temperatures.

\paragraph*{Slag viscosity formulation}

The viscosity of the slag strongly depends on the melt composition,
including the amount of constituent solid particles and the temperature
of the slag phase. The viscosity of the complete slag phase is estimated
in terms of the effective viscosity of melts containing all kinds
of particles with the Einstein-Roscoe equation. \cite{Roscoe1952}

\begin{equation}
\eta_{eff}=\eta_{m}\left(1-\frac{\phi}{\omega}\right)^{-2.5}
\label{eq:eff-visco}
\end{equation}

Where $\eta_{eff}$ is the effective viscosity of the slag phase with
a volume fraction $\phi$ of crystals, and $\eta_{m}$ is the viscosity
of the liquid part of the slag. The parameter $\omega$ corresponds
to the crystal fraction or the solid phase, at which a transition
of the system to a rigid state occurs. This parameter is obtained
from optimised experimental data. Researchers in the past have used
values between ($0.6-1$) for ranging slag systems. \cite{Liu2018,Rosenberg2005}
We have used the value of $0.6$ for our current simulations. Effective viscosity in a slag system majorly depends on the phase
fraction of the corresponding phases. The monoxide phase is responsible for the solid fraction of the slag. And thus has a certain
amount of solid associated with it. Further, this equilibrium model
was used to calculate the solid fraction which comes out to be 0.014
as shown in Table~\ref{tab:props-slag-thrmo-phy}. 

\paragraph*{Slag density formulation}

In any multiphase system, the density is found to be proportional
to the reciprocal of the molar volume of the components. In the current
work, we have implemented the density of molten slag in the following
manner as represented by Eq.~\ref{eq:rho-slag}.

\begin{equation}
\rho_{slag}=\frac{X_{i}M_{i}}{V_{m}}\label{eq:rho-slag}
\end{equation}

where $\rho$ is the density of the slags, $X_{i}$ is the mole fraction
of component $i$, $M_{i}$ is the molar mass of component $i$, and
$V_{m}$ is the molar volume of slag. The molar volume of slag was
calculated according to molar volume of pure components $V_{m}=X_{i}V_{i}\left(T\right)$
. The component molar volume $V_{i}\left(T\right)$ as shown in Table~\ref{tab:opti-Vir-dVidT}
for the various components was further expressed in form of linear
dependence of temperature in Eq.~\ref{eq:part-moler-vol}: 

\begin{equation}
V_{i}(T)=V_{i,R}+\left(T-T_{R}\right)\frac{\partial V_{i}}{\partial T}
\label{eq:part-moler-vol}
\end{equation}

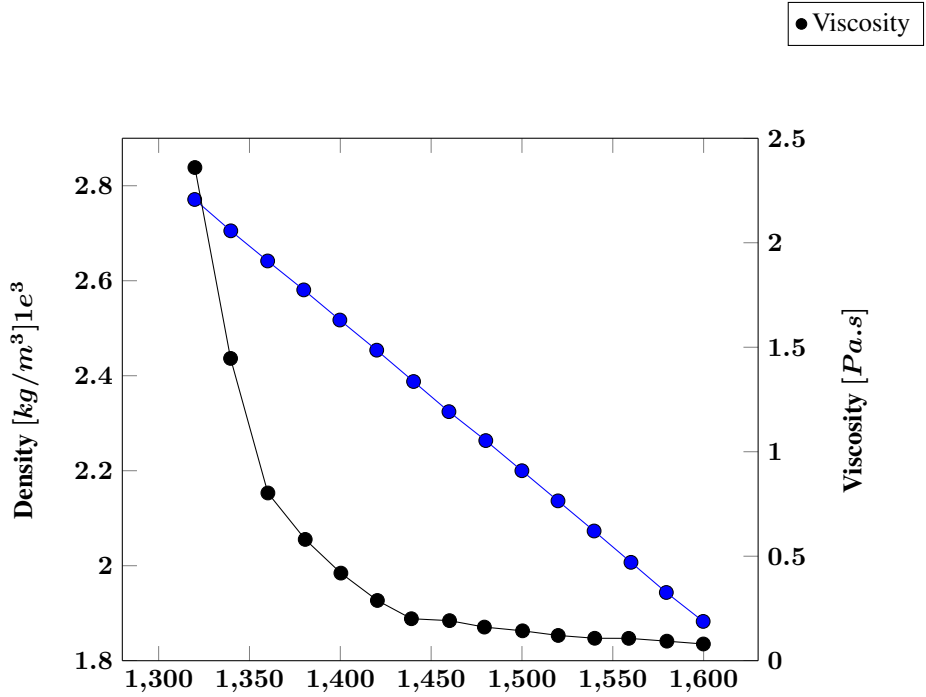
\begin{figure}[h!]
            \centering

\begin{tikzpicture}[scale=1.3]
        \pgfplotsset{width=7cm,compat=1.3}
        \pgfplotsset{
          set layers,
          mark layer=axis tick labels,
          xmin=1280, xmax=1630
        }
        \begin{axis}[
          axis y line*=left,
          ymin=1.8, ymax=2.9,
            ylabel={Density $[kg/m^3] 1e^3 $},
            label style={font=\bfseries\boldmath},
            tick label style={font=\bfseries\boldmath}
        ]
            \addplot[black,only marks,mark options={fill=blue}] table [x=Temperature, y=Density, col sep=comma]{fig6.csv};\label{plot_one}
            \addplot[blue] table [x=Temperature, y=Density, col sep=comma]{fig6.csv};\label{plot_one}
        \end{axis}
    
        \begin{axis}[
          axis y line*=right,
          axis x line=none,
          ymin=0, ymax=2.5,
          ylabel= {Viscosity $[Pa.s]$},
          label style={font=\bfseries\boldmath},
          tick label style={font=\bfseries\boldmath},
          legend pos= north east,
        ]
            \addplot[black,only marks,mark options={fill=black}] table [x=Temperature, y=Visc, col sep=comma] {fig5.csv};\label{plot_two}
            \addplot[black] table [x=Temperature, y=Visc, col sep=comma] {fig5.csv};\label{plot_two}
            \legend{Viscosity}
        \end{axis}
        
    \end{tikzpicture}

            \caption{Plot of density considering the temperature dependent Molar volume of constituent slag phases.}
            \label{fig5}
\end{figure}

Where $V_{i,R}$ is the partial molar volume of the component ‘$i$’
at reference temperature $T_{R}$, and $\partial V_{i}/\partial T$
is the thermal expansion coefficient of component ‘$i$’ as shown
in Table~\ref{tab:opti-Vir-dVidT}. The viscosity and density of slag phase for the above slag composition
in the operating slag temperature are calculated using the above Eqs.~\ref{eq:eff-visco}--\ref{eq:part-moler-vol}. The results are tabulated in the following Table~\ref{tab:props-slag-thrmo-phy}. The Effective viscosity ($\eta_{eff}$) is plotted and fitten with a least square error trendline of a polynomial function with a degree 5 exponent in the Fig.~\ref{fig5}. Similarly, for the various discrete temp, slag’s density ($\rho_{slag}$)
was calculated and then interpolated to build the UDF for density
and viscosity. As shown in the Fig.~\ref{fig5} where
slag density against temperature u relation in Eq.~\ref{eq:part-moler-vol}.

\begin{table}
\centering
\caption{Optimized parameter values of $V_{i,R}$ and $\partial V_{i}/\partial T$\cite{Glaser2012}}
\begin{tabular}{ccc}
\hline 
Components & $V_{i,R}$($1773K$)$(m^{3}mol^{-1})$ & $\partial V_{i}/\partial T$$(m^{3}mol^{-1}\,K^{-1})$\tabularnewline
\hline 
$SiO_{2}$ & $26.312\times10^{-6}$  & $0.740\times10^{-9}$ \tabularnewline
$Al_{2}O_{3}$ & $28.700\times10^{-6}$  & $10.108\times10^{-9}$ \tabularnewline
$CaO$ & $18.0312\times10^{-6}$  & $1.014\times10^{-9}$ \tabularnewline
$MgO$ & $12.076\times10^{-6}$  & $0.683\times10^{-9}$ \tabularnewline
\hline 
\end{tabular}\label{tab:opti-Vir-dVidT}
\end{table}

\begin{table}
\centering
\caption{Slag thermophysical properties of viscosity and density for the steel
refining operating temperatures}
\begin{tabular}{ccccccc}
\hline 
\multicolumn{1}{c}{Temp ( $^\circ{C}$)} & $\mu_{app}$($Pa\,s$) & Solid wt. ($g$) & Solid frac. & Ein. Richoe Coeff. & $\eta_{eff}$ ($Pa\,s$) & $\rho$$\times10^{3}$ ($kg/m^{3}$)\tabularnewline
\hline 
1600 & 0.083 & 1.389 & 0.014 & 1.036 & 0.086 & 1.883\tabularnewline
1580 & 0.091 & 1.235 & 0.012 & 1.032 & 0.094 & 1.946\tabularnewline
1560 & 0.101 & 1.8 & 0.018 & 1.046 & 0.106 & 2.010\tabularnewline
1540 & 0.112 & 2.217 & 0.022 & 1.058 & 0.118 & 2.073\tabularnewline
1520 & 0.125 & 2.54 & 0.025 & 1.066 & 0.133 & 2.136\tabularnewline
1500 & 0.139 & 2.735 & 0.027 & 1.072 & 0.149 & 2.200\tabularnewline
1480 & 0.155 & 2.973 & 0.030 & 1.078 & 0.167 & 2.263\tabularnewline
1460 & 0.174 & 3.198 & 0.032 & 1.085 & 0.189 & 2.327\tabularnewline
1440 & 0.196 & 3.412 & 0.034 & 1.091 & 0.214 & 2.390\tabularnewline
1420 & 0.221 & 10.807 & 0.108 & 1.331 & 0.294 & 2.453\tabularnewline
1400 & 0.25 & 19.469 & 0.195 & 1.718 & 0.430 & 2.517\tabularnewline
1380 & 0.283 & 25.081 & 0.251 & 2.058 & 0.583 & 2.580\tabularnewline
1360 & 0.322 & 30.775 & 0.308 & 2.508 & 0.808 & 2.643\tabularnewline
1340 & 0.368 & 42.198 & 0.422 & 3.937 & 1.449 & 2.707\tabularnewline
1320 & 0.422 & 49.798 & 0.498 & 5.600 & 2.363 & 2.770\tabularnewline
\hline 
\end{tabular}\label{tab:props-slag-thrmo-phy}
\end{table}

\paragraph*{Mass transfer in refining ladle}

The steel-slag interface is a liquid-liquid interface which is continuously
subjected to turbulence due to the continuous stirring of the steel
ladle. Various turbulence models are there which account for the continuous
renewal of fluid elements at the reaction interface and their corresponding
mass transfer rates \cite{Wilcox1988}. In this work, we have used
the surface renewal model of turbulent mass transfer theory for transfer
between two liquids \cite{Danckwerts1951}. This new theory is based
on the eddy models of turbulence and depends on parameters of turbulence,
primarily eddy-viscosity related such that it fits on some solvable
equations. Similar to the other turbulent mass transfer models, the
‘Eddy theory’ describes the mass transport of species using turbulent
eddies present at the interface of the liquids. This model provides
this link between eddy theory and mass transfer by making use of the
spectrum of turbulent energy and superimposing it with the mass transfer behaviour of an idealised small eddy motion in the vicinity of the
interface. The fluctuating turbulent
velocities are assumed to be the dominating field of velocity near
the interface.

Following the ‘small eddy model’ we need to define the mass transfer
of reacting species across the slag metal interface. For this we define
a mass transfer source term, which can be computed as follows in Eq.~\ref{eq:mass-tranf-src}.

\begin{equation}
S_{i}=m_{p^{j}q^{i}}=k_{pq}a_{i}\left(\rho_{q,e}^{j}-\rho_{q}^{j}\right)
\label{eq:mass-tranf-src}
\end{equation}

Here $S_{i}$ represent the general interphase mass transfer source
term between phase $p$ and $q$, and $(\rho_{q,e}^{j}-\rho_{q}^{j})$
is the potential to this mass transport, with $\rho_{q,e}^{j}$ being
the equilibrium mass fraction and $\rho_{q}^{j}$ being the present
mass fraction. The same equation (Eq.~\ref{eq:mass-tranf-src}) is
used in the Section~\ref{subsec:rslts-non-iso-condi} (Eq.~\ref{eq:evo-mass-trans-rate}.)
to describe the desulfurization behavior of the ladle. There we will
compute the integral of the rate of change of element sulfur due to
the interfacial reaction. Further, to relate the fluid flow and mass
transfer mathematically by accounting the rate coefficient ‘$k_{pq}$’
written as ‘$k_{L}$’ (in $m/s$) we express the following relation
in Eq.~\ref{eq:rate-coef}:

\begin{equation}
k_{L}\propto\left(\frac{\mu}{D}\right)^{-1/2}\left(\epsilon\mu\right)^{1/4}
\label{eq:rate-coef}
\end{equation}
The expression for the rate coefficient ‘$k_{L}$’ is expressed in
the above equation. Where $k_{L}$ is the mass transfer coefficient,
$\nu$ ($m^{2}/s$), is the kinematic viscosity of the metal phase,
$D$ ($m^{2}/s$) represents the Molecular diffusivity of concerned
transporting species in the metal phase and $\epsilon$ ($m^{2}/s^{3}$)
the kinetic energy dissipation rate of the metal phase by the above
lying slag phase. Using the ‘Eddy cell model’ we have the relation
between the mass transfer coefficient and turbulent flow variables.
The proportionality constant for the above relation is taken to be
$0.4$, as experimentally determined for various flows. \cite{Issa1986}
For a crude sense, a high proportionality constant would capture the
effects of larger eddies into mass transfer. In the current work,
we assume the diffusivity $D$ of Sulfur in metal is $3.5\times10^{-9}$
constant throughout the process, as the effects of non-isothermal
changes are limited in the metal phase. Thus to limit complexities
we have assumed a constant metal phase diffusivity. \cite{Mavriplis2003}

\subsection{Solution procedure}
\label{subsec:Solution-procedure}
\paragraph*{Geometry specifications}

A highly intensive geometry and meshing is used to describe and discretise
the physical geometry, such that the following criteria as well met. 
\begin{enumerate}
\item The hexahedral mesh in most parts of the geometry is chosen for better
computational efficiency, 
\item The volume of the cells for the DPM region are selected such that
particle loading is not overburdened. 
\item The slag-steel interface is finely meshed to capture the numerical
shape changes from the VOF method, such that the slag-eye formation
is appropriately addressed.
\item The zones with low fluctuations of velocity are coarsely meshed to
gain computational efficiency 
\item The rectangular porous plug region is adapted with the cylindrical
ladle such that low skewness levels are achieved overall. 
\item Finally, a multizone method is used for coupling the different zones
to achieve proper coupled solution.
\begin{figure}[h!]
\centering 
\includegraphics[width=1\linewidth]{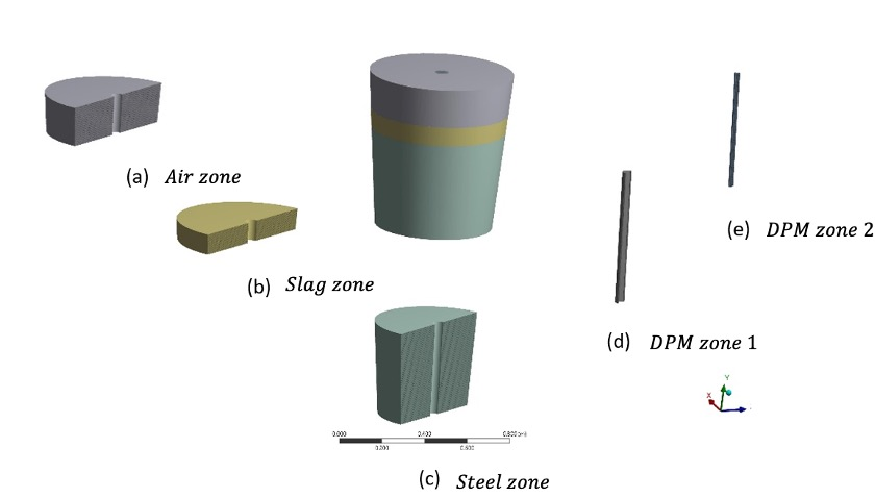} 
\caption{Physical domain converted into components of geometry. The various
domains from (a-e) are created in a manner such that an optimum balance
between the density of mesh points and features of the flow is observed
for computational efficiency.}
\label{fig:geo-dom-parts}
\end{figure}
\end{enumerate}
The geometry was created with the help of five different zones as
shown in Fig.~\ref{fig:geo-dom-parts}, The coarse Air zone (Fig.~\ref{fig:geo-dom-parts}(a))
were not much flow was expected, followed by a very fine slag zone
(Fig.~\ref{fig:geo-dom-parts}(b)) where high rates of change of
velocity vector along with rapid interfacial deformations were expected.
This was followed by a medium coarse mesh for the steel zone (Fig.~\ref{fig:geo-dom-parts}(c)),
where predominantly large scales of motion were expected in the form
of recirculation. Finally in order to efficiently capture the trajectories
of the discrete phase particles, a separate zone DPM zone 1 (Fig.~\ref{fig:geo-dom-parts}(d))
with high cell volume was created and adapted with the previous zones
using DPM zone 2 (Fig.~\ref{fig:geo-dom-parts}(e)). Fluent 18.0
Mesh building tool is used to create this customised mesh following
the above criteria. The details about the mesh quality and sizing
are described in Table~\ref{fig:mesh-struc}. The final mesh with
the given physical domains and detailed mesh structure is shown in
Fig.~\ref{fig:mesh-struc}.
\begin{figure}[h!]
\centering 
\includegraphics[width=1\linewidth]{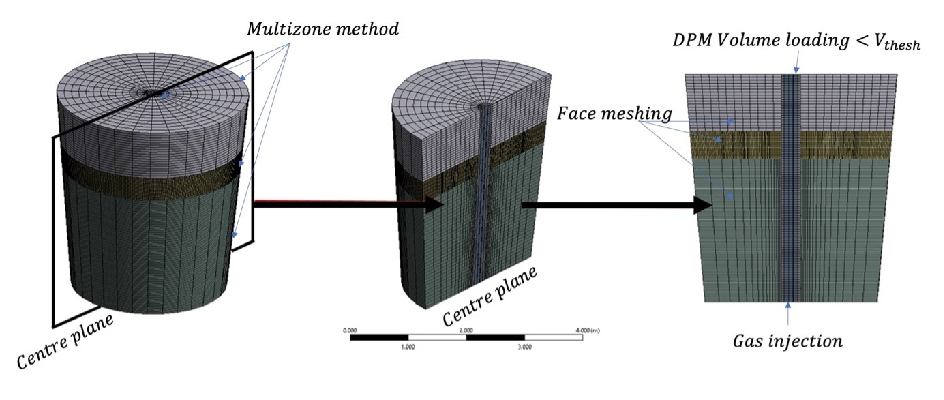}
\caption{Showing the mesh structure used as stencil to perform the domain discretisation}
\label{fig:mesh-struc}
\end{figure}

\subsection{Methodology}

The governing equations of unsteady state are to be solved on the
mesh points which are shown in the previous section. The following
numerical methods with First-order Up-wind and PISO algorithm for pressure-velocity coupling  were implemented
for obtaining the final set of algebraic equations after the governing
equations were implemented on the mesh obtained in Section~\ref{subsec:Solution-procedure}. 

\paragraph*{Solution control}

Under-relaxation factors used in the simulation are: 0.8 for discrete
phase source, 0.8 Turbulent kinetic energy, 0.8 Turbulent Dissipation
rate, 0.7 for momentum. The rest of the factors are set to 0.8.

\paragraph*{Convergence:}

The convergence for each time step for the continuum phase and discrete
phase solver is attained when the error in flow variables in successive
iterations comes below 1e-5. 

\paragraph*{Time-step control:}

Using fixed time stepping method at 0.001 sec ‘Time step size’ . The
DPM particles ejection from the inlet are set for each continuum phase
time step. 

\subsection{Solution algorithm}

\begin{figure}[h!]
\centering 
\includegraphics[width=1\linewidth]{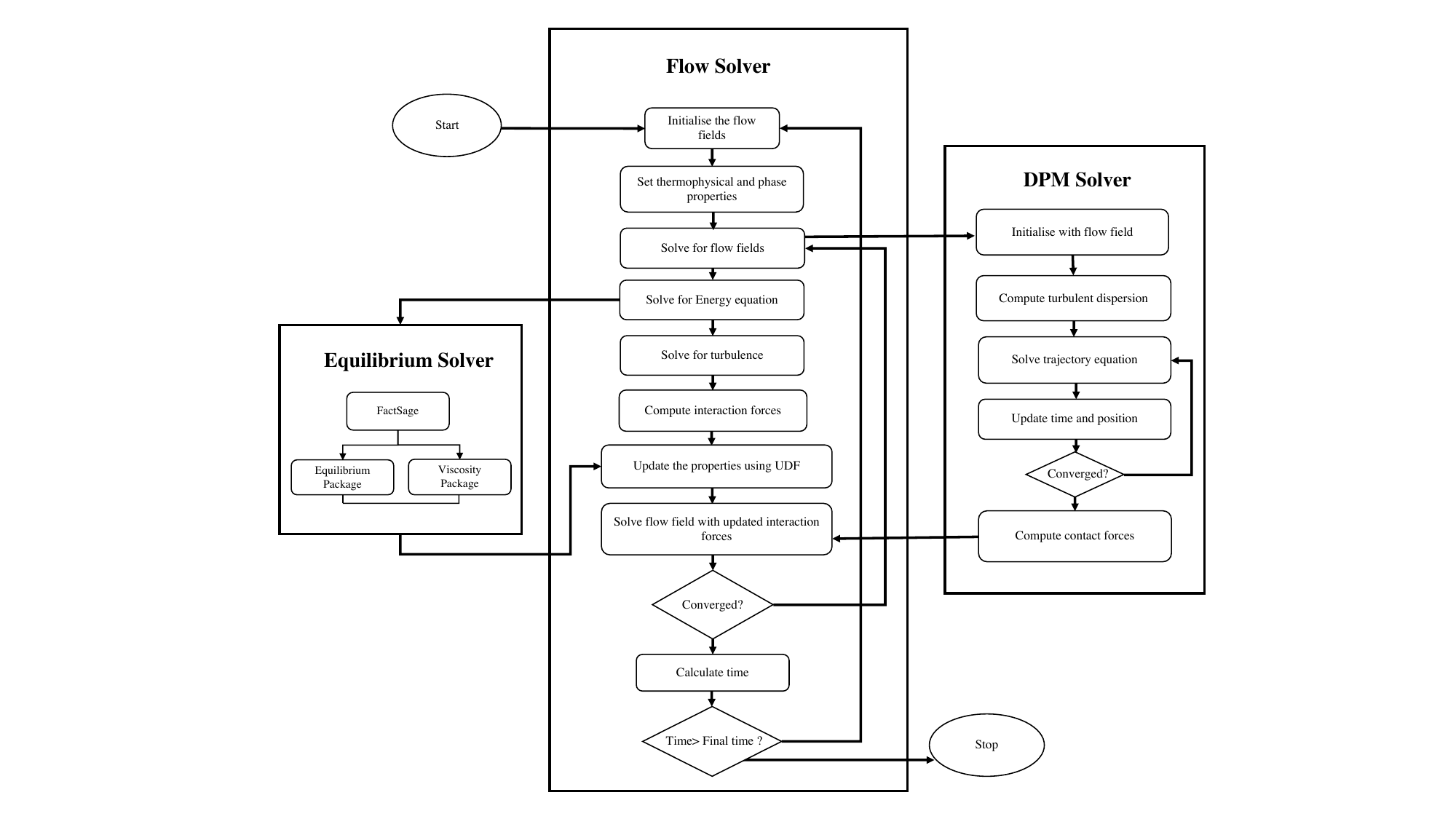}
\caption{The Solution algorithm used in the current problem.}
\label{fig:soln-algo}
\end{figure}

The given problem is solved by implementing a coupled Eulerian-Discrete
solution, where continuum phases like steel phase, slag phase and
air phase are modelled by Eulerian modelling while the discrete phases
like Argon gas bubbles (from bottom Argon gas purging) are modelled
with discrete phase modelling. The coupling is achieved with the help
of source and sink terms. The current solution is modelled with a
2-way coupling approach as shown in Fig.~\ref{fig:soln-algo}, where
the interaction forces are computed separately for both the phases
in an explicit manner.

\paragraph*{Solution algorithm }

The coupled equations of motion for incompressible isothermal multi-fluid
flow are solved using a 2 way Eulerian-DPM coupled approach. The UDF
implementation is linked with a 1 way explicit coupling with the Equilibrium
solver. The solution algorithm for advancing the solution of the flow
field from the first time-step is as follows:
\begin{enumerate}
\item Initialise the solution with the initial conditions as given in the
problem description (Section~\ref{sec:Prob-discri}). Compute new
density and viscosity distributions with the help of User defined
function and initial conditions. 
\item Compute the flow and pressure fields ($v,p$) at the discretised mesh
points using the continuity and momentum equations with the help of
the methods described in Section~\ref{subsec:Solution-procedure}.
Solve for the energy and turbulence variables $E$, $k$ and $\epsilon$
with the help of energy and turbulence equations and the flow fields
computed in the above step. 
\item Compute the interaction forces like Drag and buoyant forces and send
these forces to the discrete phase model equations as a source and
sink terms. 
\item Solve the trajectory equation for the discrete phase particles till
convergence of $10^{-5}$ for calculating variables is reached. Afterwards,
recompute the drag force. Send this updated drag force back to the
continuum flow solver. 
\item Use the updated interaction forces and recompute the continuum phase
flow variables till convergence is reached. 
\item Update the viscosity and density for the next time-step using the
user-defined functions for viscosity and density and update flow time. 
\item Return to step 2 while using the current solution for flow fields
and phase properties as initial conditions. 
\item Stop after reaching the final time. 
\end{enumerate}

\section{Validation}

The numerical model which has been presented in Section~\ref{sec:num-modeling}
has been validated with the experimentation done by previous publications
\cite{Krishnapisharody2006} Here, an acrylic container vessel was
used to simulate the slag eye formation, it was used to experiment
with a steel ladle with the ladle being scaled down to $1/10$th.
Air was injected through a central nozzle unit, which was 3mm in diameter,
The experimental ladle was of height 50 cm and the diameter was calculated
to be 42 cm. A schematic of the experimental apparatus is discussed
in the literature of the paper \cite{Krishnapisharody2006}.

\begin{figure}
        \centering
        \begin{minipage}{0.35\textwidth}
            \centering
            \includegraphics[width=1\linewidth]{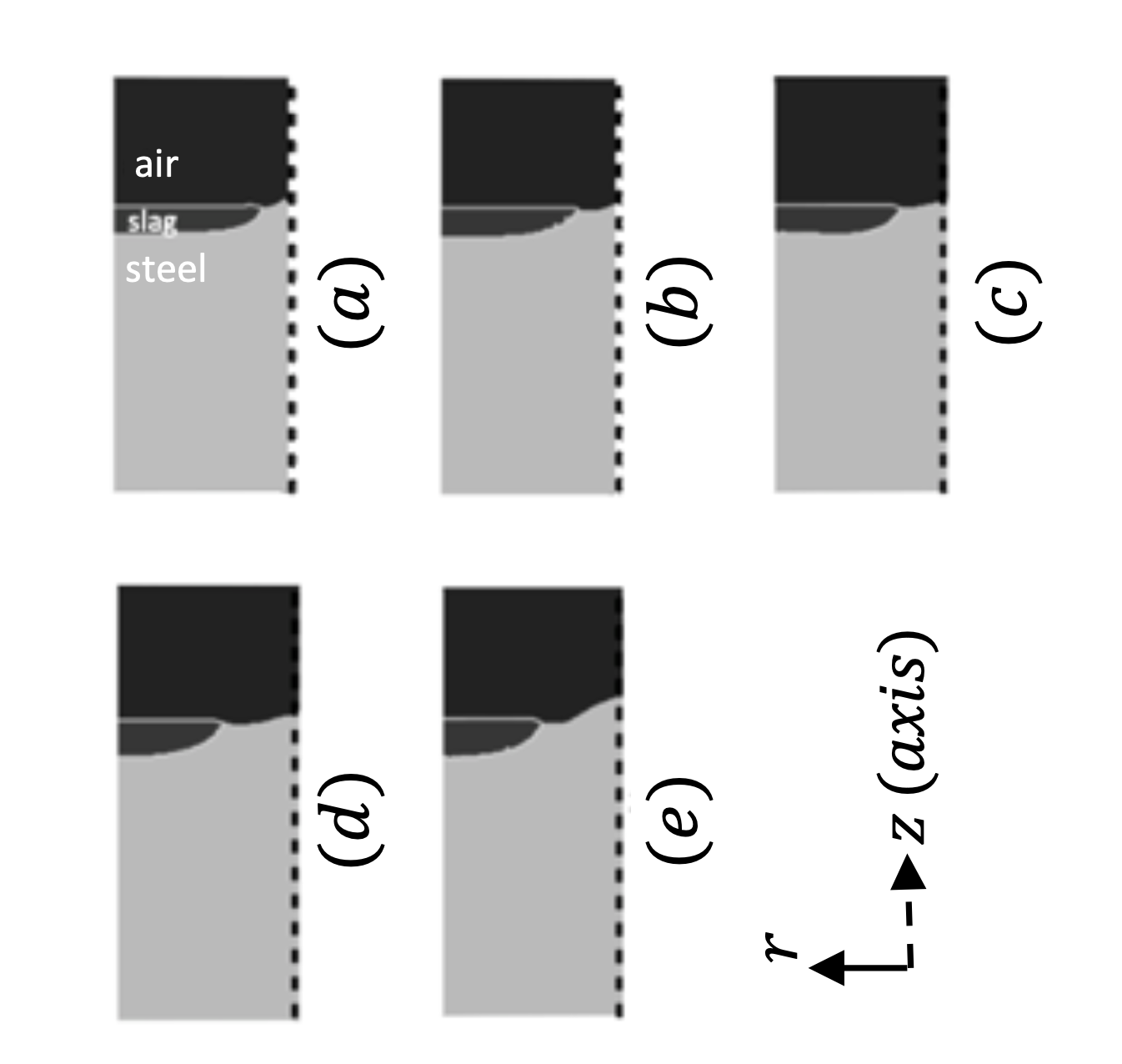}
            \caption{Validation part Effect of flow rate on eye size with (3 cm oil) for (H/D = 0.75) }
            \label{fig_l1}
        \end{minipage}\hfill
        \begin{minipage}{0.45\textwidth}
            \centering
            \pgfplotsset{every axis/.append style={scale=0.8}}


\begin{tikzpicture}
\begin{axis}[
     xmin=0, xmax=12,      
     ymin=0.15, ymax=12,  
    xlabel={Flow rate $[Lt/min]$}, 
    ylabel={Slag eye radius $[m]$},
    legend pos = north west,
    label style={font=\bfseries\boldmath},
    tick label style={font=\bfseries\boldmath}
]
\addplot+[black,only marks,mark options={fill=black}] table [x=Argon, y=Numerical, col sep=comma] {fig12.csv};
\addplot+[black,only marks,mark options={fill=blue}] table [x=Argon, y=Experimental, col sep=comma] {fig12.csv};

\legend{Experiment, Isothermal, Non-isothermal}

\end{axis}
\end{tikzpicture}
            \caption{Comparison between experimental and modelled result}
            \label{fig_l2}
        \end{minipage}
\end{figure}

The eye area for various operating conditions was obtained from video
recordings in the experiment. The eye area fluctuates over time due
to the stochastic nature of the bubble rising in the plume. The same
experiment was simulated with the existing model parameters for the
water and oil phase and the air bubbles prepared as DPM particles.
The results of the present model are consistent with those of the
experiments as shown in \cite{Krishnapisharody2006}.
However, we observe low predictability of the results at low flow
rates (as shown in Fig.~\ref{fig_l2},
which could possibly be due to the higher number of discrete phase
particles in a single volume at low flow rates, thereby decreasing
the efficiency of the solver to handle and track the particle movements
well. The ANSYS Fluent solver mentions this as a remark in the user
manual \cite{Fluent2011}, that the percentage of DPM particles in
any given cell should not exceed 15\% by weight. Otherwise, the solution
to the continuity equation of the continuous phase suffers. However,
at a more significant flow rate, the particles travel faster from
one cell to another, and thus, it becomes easier for the solver to
handle these particles.

\begin{figure}[h!]
            \centering


\begin{tikzpicture}
    \begin{axis}[
         xmin=0, xmax=300,      
         ymin=0.012, ymax=0.032,  
        xlabel={Time(sec)}, 
        ylabel={Concentration of Sulphur in metal by $[\%wt]$},
        label style={font=\bfseries\boldmath},
        tick label style={font=\bfseries\boldmath}
    ]
    \addplot+[black,only marks,mark options={fill=black}] table [x=time1, y=Experimental, col sep=comma] {fig13.csv};
    
    \addplot[scatter, no marks, draw=red] table [x=time2, y=Isothermal, col sep=comma] {fig13.csv};
    \addplot[scatter, no marks, draw=blue, dashed] table [x=time3, y=Non-isothermal, col sep=comma] {fig13.csv};
    
    \legend{Experiment, Isothermal, Non-isothermal}
    
    \end{axis}
\end{tikzpicture}
            \caption{Comparison of isothermal and non-isothermal model with respect to experimental data \cite{Jonsson1996}}
            \label{fig:comp-iso-no-iso}
\end{figure}
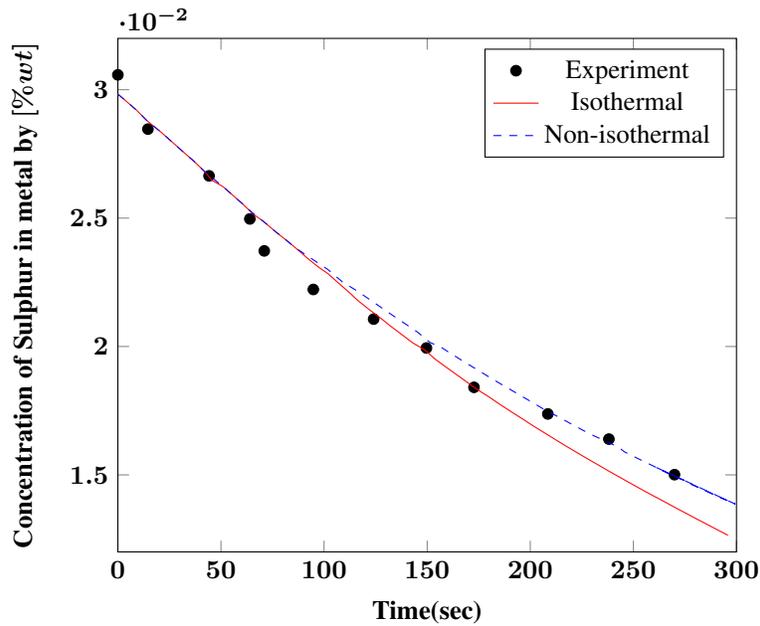

Capturing the various physics in the ladle furnace, we need to systematically
present the results to link them in sequence with the physics. 

In this part, the current numerical model for non-isothermal ladle
refining has been validated against previous experimental work from
\cite{Jonsson1996} Here we have modelled the
same experimental setup numerically and tried to get the results for
the case with the same geometrical configurations of the ladle. The
Sulfur slag metal reaction was taken to be a mass transfer phenomenon
and was demonstrated in the later Section~\ref{subsec:rslts-non-iso-condi}.
The physical parameters, and initial and boundary conditions for steel
and slag are taken the same as in the experiment. And a bath flow
rate of $80Nlm$ is implemented in a bath of $2.1m$, with the initial
Sulfur concentration being $0.031\%$$wt.
$. 

In Fig.~\ref{fig:comp-iso-no-iso} we observe the desulfurization
behaviour of both the Non-isothermal and isothermal models with respect
to the experimental measurements. The experimental results match well
with our current Non-isothermal model, while the isothermal model
tends to overpredict the results, especially in the later period of
desulfurization, where the effect of non-isothermal conditions starts
to take major effects. The study here presents the validation of this
model. Moreover, we also observe that the non-isothermal model achieves great accuracy without the need of any complex species transport or surface reaction formulation, which supports our use of mass transfer driven de-sulfurisation.

\section{Results and discussions}
\label{sec:Results-and-discussions}
The results of the current work are discussed in three separate sections,
which deal with aspects of ladle refining in a step-by-step manner. 

a) Slag-eye formation during ladle processing in the non-isothermal
modelling framework. How slag eye gets formed. b) Results related
to non-isothermal ladle conditions and how non-isothermal modelling
affects slag-eye formation c) Effects of slag eye formation on ladle
refining in non-isothermal condition.

\subsection{Slag-eye formation in steel ladle:}

\begin{figure}[h!]
\centering 
\includegraphics[width=1\linewidth]{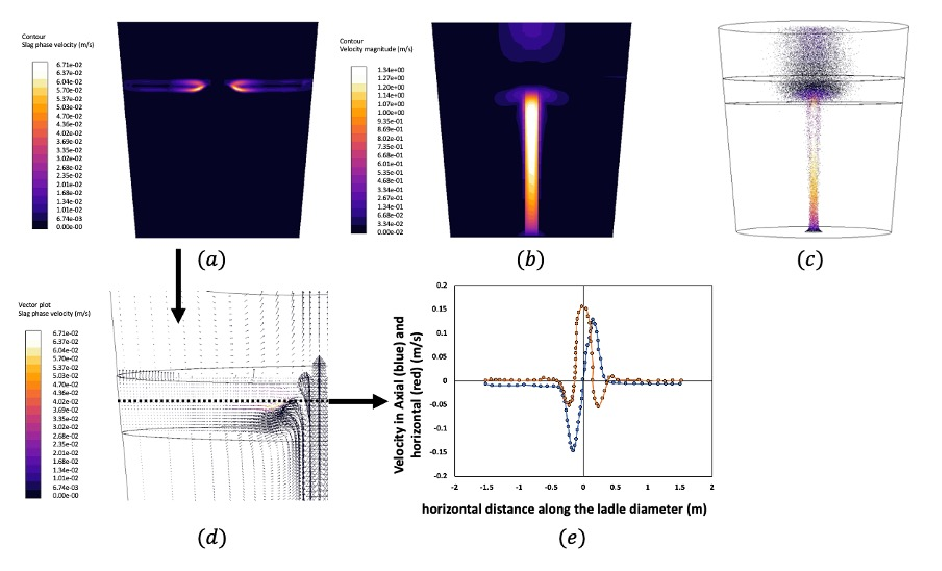} 
\caption{The contour of velocity profiles at the centre plane where (a) shows
the velocity profile in the slag phase only and (b) describes the
continuous phase mixture velocity pertaining in the ladle (c) DPM
particle in the domain marked by the velocity colourmap. Buoyancy-driven
argon bubbles are modelled as DPM particles moving in a continuous
phase. (d) The velocity vector profile in the slag region colour mapped
to the range of slag velocity. (e) A plot of axial and horizontal
velocity is shown on the ladle diameter line (the asymmetry due to
bubble-induced turbulence is observed) at ladle height = 2.45m.}
\label{fig:vel-prof}
\end{figure}

The purging of inert gas through the bottom tuyere forms a buoyant
plume of the gas-liquid mixture. The buoyant plume rises with upward
momentum. In strong purging conditions, the upward momentum of the
plume displaces the overlying slag phase and forms a slag eye-opening.
The upward-rising gas bubbles move out of the metal phase and get
into ambient air leaving the plume behind it, as shown in Fig.~\ref{fig:vel-prof}(c).
The plume then loses its upward driving force due to the downward
pull by gravitational forces and thus turns in a downward direction.
This turning of the plume also drags the adjacent slag phase and sets
up a recirculating motion in the slag phase, as shown in Fig.~\ref{fig:vel-prof}(a)
and Fig.~\ref{fig:vel-prof}(d). The velocity contour plot of the slag phase velocity as shown in Fig.~\ref{fig:vel-prof}(a)
shows the velocity profile in the slag region only. Here, we observe
that most of the flow occurs in the lower region of the slag phase
which is in contact with the metal phase as it is directly in contact
with the lower moving metal phase! Apart from that, the changing thermophysical
properties of the slag phase also favour this region to be at the
maximum velocity which is discussed in the later section.

Fig.~\ref{fig:vel-prof}(b) shows the primary phase (metal phase)
flow profile due to the two-way coupling of the discrete phase (argon
bubbles) with the primary phase. The flow profile in the ladle due
to the rising plume forms a somewhat conical shape because of the
random collisions that occur in the bubble plume, increasing the dispersion
of the bubble. And the conical shape of the plume is mainly attributed
to two factors. Firstly, stochastic turbulence in the bubble, and
secondly, the downward motion or resistance provided by the slag phase,
which deflects and spreads the rising liquid at the edges of the plume.
The bubbles (discrete phase particles) in Fig.~\ref{fig:vel-prof}(c)
then move upwards in the ladle and pull the liquid with them, while
the bubble then moves upwards through the slag; The primary phase
collides with the liquid slag phase and then recedes towards the walls.
Afterwards, the continuous phase moves downward, creating circulation
inside the ladle, as can be seen in the vector plot of the velocity
profiles in Fig.~\ref{fig:vel-prof}(d). The circulations in the
slag phase result in the asymmetric velocity plot in the horizontal
direction (radial) to the axis. The slag phase moves downwards due
to recirculation as can be observed from the axial velocity plot in
Fig.~\ref{fig:vel-prof}(e). This detailed description of the flow
field pertaining to both slag and metal phases helps in understanding
the effects which arise due to the inclusion of non-isothermal conditions
and is further discussed in the later sections of this paper. 

\subsection{Results related to Non-isothermal ladle conditions}
\label{subsec:rslts-non-iso-condi}

\subsubsection*{Effect of Non-isothermal ladle conditions on slag eye formation}

At the onset of ladle processing, Argon purging is performed at varying
rates depending upon the timeline pertaining to the ladle processing,
such as a low purging rate for mixing, a high rate of purging for
de-S, etc. The argon purging brings about the effect of relatively
cooler gas being blown in the hot metal, resulting in inevitable temperature
loss. Moreover, as discussed in the previous section the upward-rising
buoyant plume also causes the creation of slag eye opening during
high purging. This way the hot metal is exposed to the ambient air
at cooler temperatures, and heat losses start to occur from this open
eye. 
\begin{figure}[h!]
\centering
\includegraphics[width=1\linewidth]{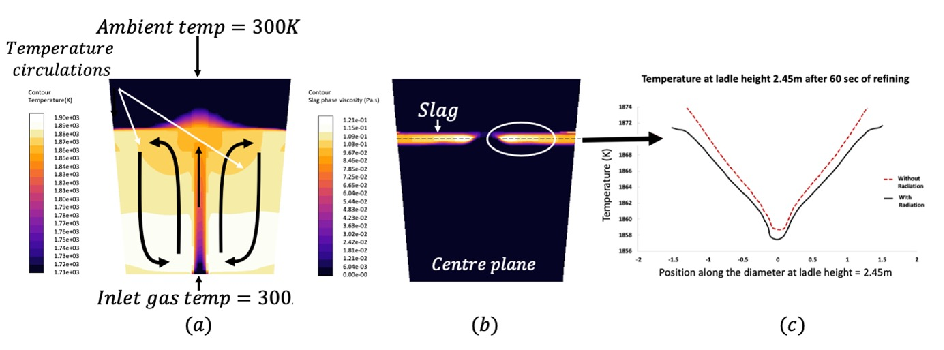} 
\caption{(a) The temperature distribution of the mixture phase in the ladle
(clipped to range 1900K-1700K), the temperature distribution contour
is after 60 sec from the start of the simulation.(b) Slag viscosity
variation in the slag eye region and (c) the plot of the corresponding
temperature at ladle height = 2.45m marked by the dashed line after
60 sec from the start of the simulation.}
\label{fig:temp-distri}
\end{figure}

We start our discussion with the results for the total temperature
profile within the ladle, as we can see from Fig.~\ref{fig:temp-distri}(a).
The coldest region exists at the ladle bottom due to cool argon purging,
and next up we see the region, which is near the open slag eye. It
can also be judged from the implications of the radiative flux model,
as the element's radiative flux contribution to another region in
the molten metal is almost uniform and thereby can’t cancel out each
other, while the elements near the slag open eye region have less
radiant flux coming from the elements from the top (i.e., the ambient)
than what it is emitting from itself. The effect of including this
radiation model can be seen in Fig.~\ref{fig:temp-distri}(c), where
the temperature profile comes out to be lower than the model with
a no-radiation model. 

The effect of the temperature is reflected upon the thermo-physical
properties of the materials and in turn the flow conditions in the
ladle. As shown in Fig.~\ref{fig:temp-distri}(b) the viscosity of
the slag phase is varying in the ladle as the temperature continuously
drops in the ladle due to the reasons described in the previous paragraph.
We see a marked viscosity increase in the region near the slag eye
region, this can be well related to the temperature profile on the
line in the inter-slag region. This viscosity increase can be well
attributed to the change in the flow profile in the slag and the interfacial
region in the metal phase. The results of these changes will be attributed
in the next sections where we will relate the flow change effects
to the macroscopic effects of refining. 

Slag eye area as observed in Fig.~\ref{fig:slag-eye-build-up}(a),
is one of the most important parameters in ladle processing as discussed
in the previous section. The thermophysical properties and the flow
regime in the interfacial region particularly near the bubble plume
are important for the dimension of the slag eye area. Fig.~\ref{fig:slag-eye-build-up}(b)
and Fig.~\ref{fig:slag-eye-build-up}(c) shows the effect of including
the variations in the thermophysical properties on the slag eye. Here
we observe a smaller slag eye area when the thermophysical properties
are considered. 

\begin{figure}[h!]
\centering 
\includegraphics[width=1\linewidth]{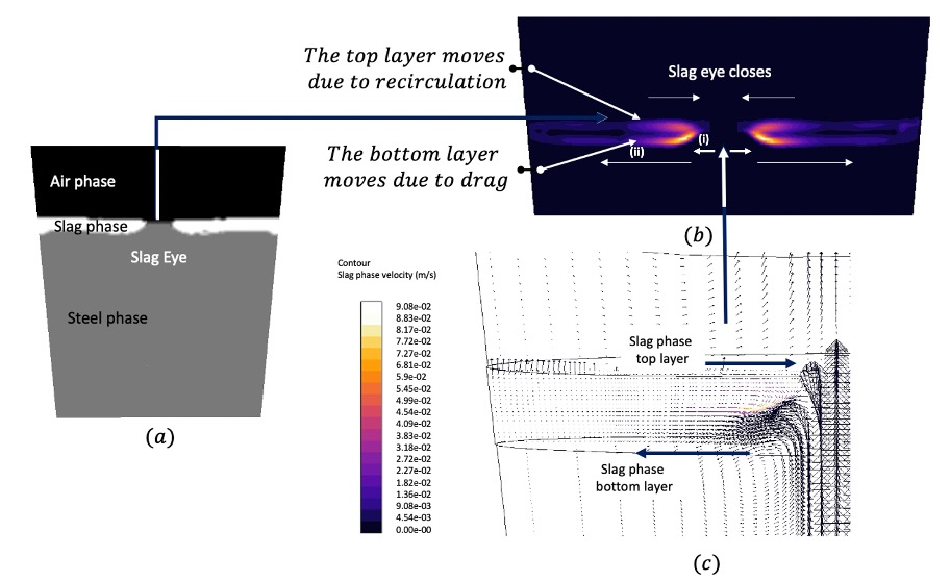} 
\caption{(a) Slag eye build-up during ladle refining operation at an argon
flow rate of 400 Lt/min. (b) Process of slag eye closure due to increasing
recirculation in the ladle slag colour mapped to slag’s velocity magnitude
(c) Velocity vector plot of continuous phases (Steel, slag and air)
colour mapped to slag’s velocity magnitude.}
\label{fig:slag-eye-build-up}
\end{figure}

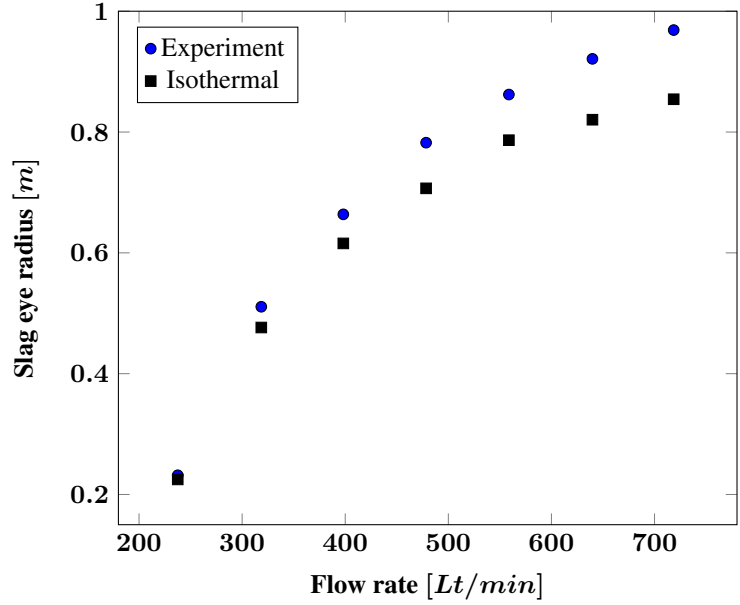
\begin{figure}[h!]
            \centering


\begin{tikzpicture}
\begin{axis}[
     xmin=180, xmax=780,      
     ymin=0.15, ymax=1,  
    xlabel={Flow rate $[Lt/min]$}, 
    ylabel={Slag eye radius $[m]$},
    legend pos = north west,
    label style={font=\bfseries\boldmath},
    tick label style={font=\bfseries\boldmath}
]
\addplot+[black,only marks,mark options={fill=blue}] table [x=Argon, y=Iso, col sep=comma] {fig17.csv};
\addplot+[black,only marks,mark options={fill=black}] table [x=Argon, y=Noniso, col sep=comma] {fig17.csv};

\legend{Experiment, Isothermal, Non-isothermal}

\end{axis}
\end{tikzpicture}
            \caption{Comparison between normal slag eye area and changing viscosity slag eye area.}
            \label{fig:slag-eye-build-up}
\end{figure}

In order to understand the effects of changing thermo-physical properties
on the flow behavior of slag. We observe Fig.~\ref{fig:slag-eye-build-up}(c),
where we have a vector plot of a zoomed ladle in the slag-metal interfacial
region. The plot is colour-mapped for the slag phase velocity magnitude.
Here we observe an accelerated (changing velocity vector direction)
slag phase as a result of the convective drag produced by the adjacent
flowing metal phase. Adding to this result the information coming
from the thermophysical plot of the slag, we see that the slag region
with maximum slag viscosity corresponds to the maximum drag produced
in the slag phase and the maximum velocity profile in the slag phase
from now on. Here we see that the slope of the slag eye increases,
but due to the higher returning fluid due to the increased recirculation,
the top part of the eye also closes, and a smaller eye is formed.
Furthermore, the slag phase in the vicinity of the open eye is relatively
denser than the rest of the slag phase. This creates a gravitational
push on the lower region of the slag phase in that region, and accordingly
the circulation in the phase which we can see in the figure.

Due to this the circulations in the slag phase increases and an increased
circulation then diminishes the slag eye region. And thus a relatively
smaller slag-eye is formed, as shown in the Fig.~\ref{fig:comp-slag-eye-area}.
Where a comparison has been shown between sag eye area made by Isothermal
and non-isothermal conditions. We observe that the mismatch in the
slag eye radius is more prominent at higher purging rates, since at
higher purging rates we form a larger slag eye and hence there is
a much larger heat loss. This larger heat loss creates a larger temperature
drop in the slag phase and thus makes it more viscous and dense. This
way the slag eye tends to close to a greater extent as explained by
the mechanism discussed in this section. 

\subsubsection*{Effects of slag eye formation in non-isothermal conditions on ladle
refining}

\begin{figure}[h!]
\centering 
\includegraphics[width=1\linewidth]{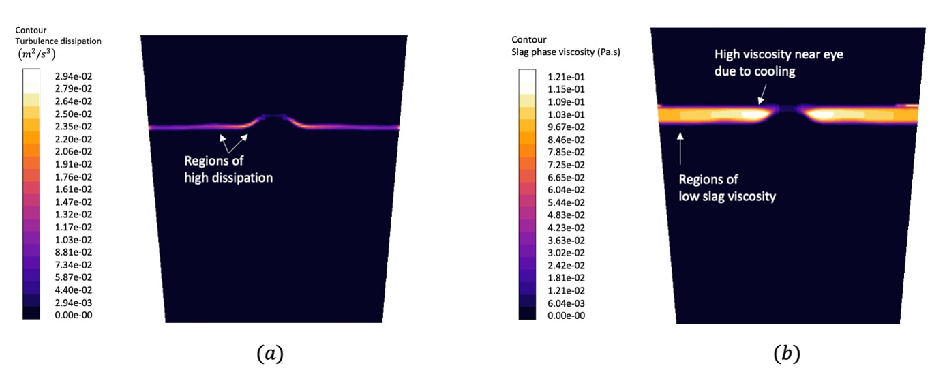} 
\caption{: (a) The contour plot of turbulence energy dissipation at the slag
metal interface, the contour plane is placed interesting the axis
of symmetry. (b) The contour plot of slag viscosity in the ladle.}
\label{fig:turb-ke-diss}
\end{figure}

The evolving thermophysical properties of the slag phase due to non-isothermal
flow conditions lead to several phenomena related to the changing
flow field conditions as discussed in previous sections. In this section,
we attempt to analyze its effects on ladle processing. Since the basic
purpose of ladle furnaces is to achieve the desired composition values
in a given processing time and to facilitate downstream processes
such as tundish metallurgy and casting, it is thus important to analyze
the impact of this study on interfacial refining reactions. In order
to study the effects on interfacial refining reactions, we have used
the small eddy mass transfer formulation following the surface renewal
principle suggested by Lamont \cite{Lamont1970}. According to this
theory dissipation in the interfacial region due to turbulence in
the flow, the field is the main major factor in the mass transfer
between the two fluid phases. As shown in Fig.~\ref{fig:turb-ke-diss}(a),
we analyze the role of changes in thermophysical properties on changes
in disturbance dissipation energy, and further, we explain the steel
phase convection mass transfer from this dissipation energy according
to Eq.~\ref{eq:mass-tranf-src}., outlined in D6 is correlated.

From Fig.~\ref{fig:turb-ke-diss}(a) and Fig.~\ref{fig:turb-ke-diss}(b)
we observe that turbulence energy dissipation is enhanced in the region
where the slag’s viscosity is lower. Highly viscous slag requires
more energy to displace and thus less amount of intermixing is observed
in these regions. This is further explained in the context of ladle
refining with the help of de-desulfurization of the metal phase. For
this, we have defined the turbulent mass transfer of Sulfur between
the steel and slag phase at the steel-slag interface. The overall
mass transfer coefficient for this process is computed based on the
Integration of the mass transfer coefficient expression at the interface
as shown in Eq.~\ref{eq:mass-tran-coef}:

\begin{equation}
k_{l}=\int_{\underset{f\rightarrow5}{lim}}\alpha_{steel\pm f\%}\,\alpha_{slag\pm f\%}\left(\frac{\mu}{D}\right)^{-1/2}\left(\epsilon\mu\right)^{1/4}dV
\label{eq:mass-tran-coef}
\end{equation}

In Fig.~\ref{fig:turb-ke-diss}(b) we have the viscosity of the slag
phase, and we see that the interfacial region has a relatively high
viscosity due to the cooling and circulation effect of the slag phase.
However, this interface region is constantly exposed to high-temperature
molten steel. This high viscosity slag produces high turbulence dissipation
of the steel phase turbulent kinetic energy in this region but due
to the high slag viscosity, the intermixing in this zone becomes sluggish.
As a result, to that, we observe a relatively low slag-metal interfacial
area, and thus a decrement in mass transfer coefficient is observed.
These results are taken at a fixed sample of time, further, we present
the results obtained from transient analysis where non-isothermal
cooling of the ladle is taking place. Here, we have presented the
evolution of the mass transfer coefficient and turbulent energy dissipation
in the lava phase as a function of ladle processing time.

\begin{equation}
\left[wt\%S\left(t\right)\right]=\int_{De-S\,time}\int_{\underset{f\rightarrow5}{lim}}\alpha_{steel\pm f\%}\,\alpha_{slag\pm f\%}K_{s}\frac{A}{V}\left(\left[wt\%S\left(t\right)\right]-\frac{\left[wt\%S\left(t\right)\right]}{L_{s}}\right)dVdt
\label{eq:evo-mass-trans-rate}
\end{equation}

The metal phase Sulfur content is predicted based on the above Eq.~\ref{eq:evo-mass-trans-rate}.
Integrating the above equation at the interface we determine the rate
of change of the bath’s sulfur content and further integrating this
rate for the de-sulfurisation time we get the time-dependent sulfur
content. In the above equation $L_{s}$ represents the equilibrium
partition coefficient. In our work, we assume $L_{s}$ to be constant
and equal to 350 for the ladle refining conditions. $K_{s}$ ($m/s$)
represents the mass transfer coefficient for the partitioning of sulfur
from metal to slag phase, as derived in the previous section. $A/V$
($m^{2}/m^{3}$) represents the interfacial area concentration of
the interacting slag-metal phases. With the above relation, we can
observe the effects of refining with changing argon purging rates.

\begin{figure}
        \centering
        \begin{minipage}{0.5\textwidth}
            \centering
            \pgfplotsset{every axis/.append style={scale=0.8}}
            \begin{tikzpicture}
\begin{axis}[
     xmin=0, xmax=350,      
     ymin=0.00035, ymax=0.00105,  
    xlabel={Time(sec)}, 
    ylabel={Turbulent dissipation rate $[m^2/s^3]$},
    legend pos= north west,
    label style={font=\bfseries\boldmath},
    tick label style={font=\bfseries\boldmath},
]
\addplot+[black,only marks,mark options={fill=black}] table [x=time, y=diss240, col sep=comma] {fig18.csv};
\addplot[black,only marks,mark options={fill=blue}] table [x=time, y=diss400, col sep=comma] {fig18.csv};
\addplot[black,only marks,mark options={fill=red}] table [x=time, y= diss560, col sep=comma] {fig18.csv};
\addplot[black,only marks,mark options={fill=pink}] table [x=time, y= diss720, col sep=comma] {fig18.csv};

\legend{240 N/l, 400 N/l, 560 N/l, 720 N/l}

\end{axis}
\end{tikzpicture}
            \caption{Comparsion of dissipation rate }
            \label{fig18a}
        \end{minipage}\hfill
        \begin{minipage}{0.5\textwidth}
            \centering
            \pgfplotsset{every axis/.append style={scale=0.8}}
            \begin{tikzpicture}
\begin{axis}[
     xmin=0, xmax=350,      
     ymin=1e-6, ymax=5e-6,  
    xlabel={Time(sec)}, 
    ylabel={Mass transfer rate $[m/s]$},
    label style={font=\bfseries\boldmath},
    tick label style={font=\bfseries\boldmath},
][legend pos=outer north west]
\addplot+[black,only marks,mark options={fill=black}] table [x=time, y=ms240, col sep=comma] {fig19.csv};
\addplot[black,only marks,mark options={fill=blue}] table [x=time, y=ms400, col sep=comma] {fig19.csv};
\addplot[black,only marks,mark options={fill=red}] table [x=time, y= ms560, col sep=comma] {fig19.csv};
\addplot[black,only marks,mark options={fill=pink}] table [x=time, y= ms720, col sep=comma] {fig19.csv};

\legend{240 N/l, 400 N/l, 560 N/l, 720 N/l}

\end{axis}
\end{tikzpicture} 
            \caption{Comparsion of mass transfer rate }
            \label{fig18b}
        \end{minipage}
\end{figure}

As seen in Fig.~\ref{fig:turb-ke-diss}(a), we observe that the increasing
viscosity increases the dissipation of energy near the interface.
As shown in Fig.~\ref{fig18a} the argon
gas flow rate increases more dissipation can take place near the surface
due to increased turbulence. Moreover, with an increasing argon gas
flow rate, we expect the slag cooling to take place at a higher rate
thus increasing the dissipation. With an increasing dissipation rate,
we expect the mass transfer to enhance as per Eq.~\ref{fig18b}
but we observe some discrepancies with this inference. 

The plot of the change in the mass transfer coefficient over time
shows that the mass transfer decreases as ladle processing progresses
as seen in Fig.~\ref{fig18b}. The possible
explanation for this is attributed to the decrease in the magnitude
of the flow in the vicinity of the slag layer due to the increase
in the viscosity of the slag layer. Due to the increasing viscosity,
more energy is now required to displace the slag layer for slag-metal
intermixing and henceforth although the turbulent kinetic energy of
dissipation increases, still the mass transfer coefficient decreases
in this interfacial region. If such non-isothermal transformations
are not considered, the mass transfer and turbulence dissipation turn
out to be a constant value when averaged over time and integrated
into space. As the thermophysical slag layer will no longer change
with the progress of the refining process. Furthermore, the mass transfer
rate tends to decrease at a higher rate at higher argon purging rates.
This can be attributed to the higher rate of cooling which can be
observed at higher argon purging rates, due to the larger slag eye
area. As the slag phases cool down, the higher viscous slag tends
to decrease the intermixing of the phases at the interface. This way
the refining process will become sluggish.

\begin{figure}[h!]
            \centering


\begin{tikzpicture}
\begin{axis}[
     xmin=0, xmax=300,      
     ymin=0.005, ymax=0.032,  
    xlabel={Flow rate $[Lt/min]$}, 
    ylabel={Slag eye radius $[m]$},
    legend pos = outer north east,
    label style={font=\bfseries\boldmath},
    tick label style={font=\bfseries\boldmath}
]

\addplot[orange] table [x=Time, y=Iso240, col sep=comma] {fig20.csv};
\addplot[orange,dashed,mark options={circle,fill=yellow}] table [x=Time, y=Non240, col sep=comma] {fig20.csv};
\addplot[blue] table [x=Time, y=Iso400, col sep=comma] {fig20.csv};
\addplot[blue,dashed,mark options={circle,fill=yellow}] table [x=Time, y=Non400, col sep=comma] {fig20.csv};
\addplot[violet] table [x=Time, y=Iso560, col sep=comma] {fig20.csv};
\addplot[violet,dashed,mark options={circle,fill=yellow}] table [x=Time, y=Non560, col sep=comma] {fig20.csv};
\addplot[black] table [x=Time, y=Iso720, col sep=comma] {fig20.csv};
\addplot[black,dashed,mark options={circle,fill=yellow}] table [x=Time, y=Non720, col sep=comma] {fig20.csv};

\addplot[orange,only marks,mark options={fill=orange}] table [x=Time, y=Iso240, col sep=comma] {fig20.csv};
\addplot[orange,only marks,mark options={fill=white}] table [x=Time, y=Non240, col sep=comma] {fig20.csv};
\addplot[blue,only marks,mark options={fill=blue}] table [x=Time, y=Iso400, col sep=comma] {fig20.csv};
\addplot[blue,only marks,mark options={fill=white}] table [x=Time, y=Non400, col sep=comma] {fig20.csv};
\addplot[violet,only marks,mark options={fill=violet}] table [x=Time, y=Iso560, col sep=comma] {fig20.csv};
\addplot[violet,only marks,mark options={fill=white}] table [x=Time, y=Non560, col sep=comma] {fig20.csv};
\addplot[black,only marks,mark options={fill=black}] table [x=Time, y=Iso720, col sep=comma] {fig20.csv};
\addplot[black,only marks,mark options={fill=white}] table [x=Time, y=Non720, col sep=comma] {fig20.csv};

\legend{Iso 720, Non-iso 720, Iso 560, Non-iso 560, Iso 400, Non-iso 400, Iso 240, Non-iso 240}

\end{axis}
\end{tikzpicture}
            \caption{Comparison of the weight percentage content of Sulfur in the molten
                    steel with respect to time during the ladle refining operation in
                    Isothermal and Non-isothermal conditions. }
            \label{fig:comp-wt-pct-content}
\end{figure}
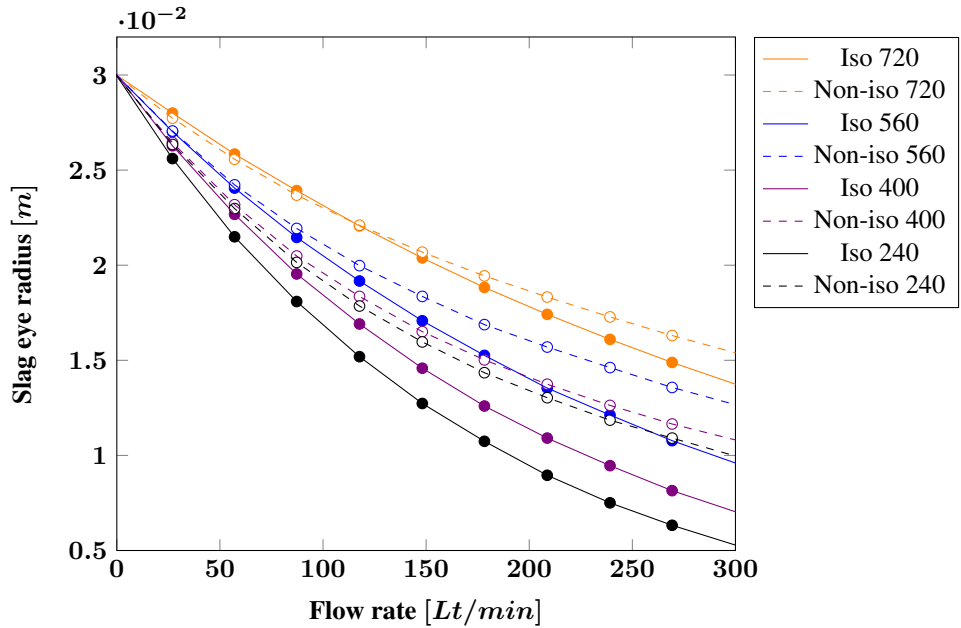

As discussed in the previous section, the non-isothermal conditions
during the ladle refining process affect the mass transfer rates of
the refining process. Assuming desulfurization to be metal phase mass
transfer controlled, and using the formulation as shown in Eq.~\ref{eq:evo-mass-trans-rate}
we solve for the metal phase Sulfur content$\left[\%S\right]$ with
respect to time. From the previous sections, a diminished mass transfer
rate leads to a lesser extent of desulfurization from the hot metal. 

From Fig.~\ref{fig:comp-wt-pct-content}, we infer that the final
sulfur content decreases as we increase the input argon flow rate.
This can be seen due to the increased mass transfer rate at higher
argon gas purging rates, as can be seen in Fig.~\ref{fig18b}.
The decrease in final Sulfur content, $\triangle S$ $\left[wt.\ensuremath{\%}\right]$
between isothermal and non-isothermal models, increases with increasing
purging rate increases as we increase the argon flow rate. However,
the change in $\left[\%S\right]$ content between isothermal and non-isothermal
conditions increases with time but with a decreasing rate with respect
to the argon gas flow rate. This effect can also be inferred from
Fig.~\ref{fig18b}, as the rate of decrement
of mass transfer rate increases with increasing purging rates. 

\section{Conclusion}

In this work, a sophisticated multiphase numerical model is implemented on Ansys Fluent. The model involves transient equations of mass, momentum, and energy under turbulent conditions. The discrete particle method is employed to handle the bubble dynamics. Volume of fluid method is employed to track the interface and the radiation is modelled to account for the heat loss from the top surface. The model can study the slag eye formation, intermixing, and refining under non-isothermal conditions. The major conclusions of this work are:

\begin{enumerate}
    \item The ladle eye opening through which high radiative heat transfer takes place along with the convective cooling, cools down the content of the ladle. Cooling of metal/slag also takes place due to radiation loss from the top and rising gas bubbles. All these results in the change of thermophysical properties, mainly the viscosity of slag.
    \item A relatively smaller slag-eye is formed in non-isothermal conditions as compared to the isothermal conditions. The mismatch is more prominent at higher purging rates as a larger slag eye produces a larger temperature drop in the slag phase and thus makes it more viscous and denser. This way the slag eye tends to close to a greater extent.
    \item The results of the slag’s kinematic study show the intersection of the high-velocity slag region with higher viscosity region which slows down the slag phase and restricts the bubble plume to grow in radius at the top. This way the slag phase provides greater resistance to slag-eye build-up by the bubble plume.
    \item A high argon purging rate corresponds to high turbulent dissipation which corresponds to higher mass transfer rates according to the eddy-dissipation model. But it is also accompanied by a higher degree of slag cooling and increasing viscosity of the slag phase. The net effect of both of these competing effects on mass transfer is observed in the model. Finally, it is concluded that the non-isothermal viscosity changes dominate and thus make the overall mass transfer sluggish.
    \item The effects of changing thermophysical properties on ladle refining are concluded by formulating its effect on the mass transfer coefficient of the steel melt. The mass transfer coefficient becomes sluggish as the refining proceeds forward. This decreasing mass transfer rate affects the desulfurization process. Thus to achieve the same level of desulphurization, a non-isothermal case would require more time or higher purging rate compared to an isothermal case.
    \item The mass transfer based desulfurisation formulation along with non-isothermal multiphase flow formulation is able to accurately predict metal refining process such as desulphurisation. This validated approach eliminates the necessity of solving computationally intensive equations related to species transport for achieving the same objective.
\end{enumerate}

The work opens up possibilities towards better and more realistic ladle refining optimization. Questions such as optimum argon flow rates in view of competing processing time and unwarranted pick-up of impurities through slag-eye can be effectively addressed. Questions on the extent and need of intermittent arcing during high purging processes like de-sulphurization can also be addressed with the current work.

\bibliographystyle{unsrt}
\bibliography{refs}

\begin{thebibliography}{10}

\bibitem{Zhang2003}
Lifeng Zhang and Brian~G. Thomas.
\newblock State of the art in evaluation and control of steel cleanliness.
\newblock {\em ISIJ International}, 43(3):271--291, 2003.

\bibitem{Ghosh2000}
A.~Ghosh.
\newblock {\em Secondary Steelmaking: Principles and Applications (1st ed.)}.
\newblock CRC Press, 2000.

\bibitem{Xie1992}
Yongkun Xie and Franz Oeters.
\newblock Experimental studies on the flow velocity of molten metals in a ladle
  model at centric gas blowing.
\newblock {\em Steel Research}, 63(3):93--104, 1992.

\bibitem{Yonezawa1999}
Kimitoshi Yonezawa and Klaus Schwerdtfeger.
\newblock Spout eyes formed by an emerging gas plume at the surface of a
  slag-covered metal melt.
\newblock {\em Metallurgical and Materials Transactions B}, 30:411--418, 1999.

\bibitem{Mazumdar2004}
D.~Mazumdar and J.W. Evans.
\newblock A model for estimating exposed plume eye area in steel refining
  ladles covered with thin slag.
\newblock {\em Metallurgical and Materials Transactions B}, 35:400--404, 2004.

\bibitem{Krishnakumar2008}
Krishnapisharody Krishnakumar and Gordon~A. Irons.
\newblock An extended model for slag eye size in ladle metallurgy.
\newblock {\em ISIJ International}, 48(12):1807--1809, 2008.

\bibitem{Wu2010}
L.~Wu, P.~Valentin, and D.~Sichen.
\newblock Study of open eye formation in an argon stirred ladle.
\newblock {\em steel research international}, 81(7):508--515, 2010.

\bibitem{Krishnapisharody2006}
K.~Krishnapisharody and G.A. Irons.
\newblock Modeling of slag eye formation over a metal bath due to gas bubbling.
\newblock {\em Metall Mater Trans B}, 37:763--772, 2006.

\bibitem{Liu2017}
Z.~Liu, L.~Li, and B.~Li.
\newblock Modeling of gas-steel-slag three-phase flow in ladle metallurgy: Part
  1. physical modeling.
\newblock {\em ISIS International}, 57(11):1971--1979, 2017.

\bibitem{Volkova2003}
Olena Volkova and Dieter Janke.
\newblock Modelling of temperature distribution in refractory ladle lining for
  steelmaking.
\newblock {\em ISIJ international}, 43(8):1185--1190, 2003.

\bibitem{Camdali2006}
Ünal Çamdali and Murat Tunc.
\newblock Steady state heat transfer of ladle furnace during steel production
  process.
\newblock {\em Journal of Iron and Steel Research, International},
  13(3):18--25, 2006.

\bibitem{Zimmer2008}
André Zimmer, Álvaro Niedersberg Correia~Lima, Rafael~Mello Trommer,
  Saulo~Roca Bragança, and Carlos~Pérez Bergmann.
\newblock Heat transfer in steelmaking ladle.
\newblock {\em Journal of Iron and Steel Research, International},
  15(3):11--60, 2008.

\bibitem{Livshits2011}
D.~A. Livshits et~al.
\newblock Heat-loss calculation in ladle treatment of steel.
\newblock {\em Steel in Translation}, 40(11):979--982, 2011.

\bibitem{Glaser2011}
Björn Glaser, Mårten Görnerup, and Du~Sichen.
\newblock Thermal modelling of the ladle preheating process.
\newblock {\em steel research international}, 82(12):1425--1434, 2011.

\bibitem{Kabakov2013}
Z.~K. Kabakov and M.~A. Pakholkova.
\newblock Reducing the loss of heat from steel in steel-pouring ladles.
\newblock {\em Metallurgist}, 56(9):670--671, 2013.

\bibitem{Seshadri2016}
Varadarajan Seshadri, Izabela~Diniz Duarte, Itavahn~Alves Da~Silva, and
  Carlos~Antonio Da~Silva.
\newblock {\em Evaluation of Heat Flow and Thermal Stratification in A
  Steelmaking Ladle Through Mathematical Modelling}, chapter~60, pages
  487--494.
\newblock John Wiley \& Sons, Ltd, 2016.

\bibitem{FarreraBuenrostro2019}
J.~E. Farrera-Buenrostro et~al.
\newblock Analysis of temperature losses of the liquid steel in a ladle furnace
  during desulfurization stage.
\newblock {\em Transactions of the Indian Institute of Metals}, 72(4):899--909,
  2019.

\bibitem{Gonzalez2017}
H.~Gonzalez et~al.
\newblock Multiphase modeling of fluid dynamic in ladle steel operations under
  non-isothermal conditions.
\newblock {\em Journal of Iron and Steel Research International},
  24(9):888--900, 2017.

\bibitem{Han2001}
Jeong~Whan Han et~al.
\newblock Transient fluid flow phenomena in a gas stirred liquid bath with top
  oil layer—approach by numerical simulation and water model experiments.
\newblock {\em ISIJ international}, 41(10):1165--1172, 2001.

\bibitem{TafaghodiKhajavi2010}
Leili Tafaghodi~Khajavi and Mansoor Barati.
\newblock Liquid mixing in thick-slag-covered metallurgical baths—blending of
  bath.
\newblock {\em Metallurgical and Materials Transactions B}, 41(1):86--93, 2010.

\bibitem{Liu2018}
W.~Liu, H.~Tang, S.~Yang, et~al.
\newblock Numerical simulation of slag eye formation and slag entrapment in a
  bottom-blown argon-stirred ladle.
\newblock {\em Metall. Mater. Trans. B}, 49:2681--2691, 2018.

\bibitem{Guo2002}
D.~Guo and G.~A. Irons.
\newblock A water model and numerical study of the spout height in a
  gas-stirred vessel.
\newblock {\em Metallurgical and Materials Transactions B}, 33(3):377--384,
  2002.

\bibitem{Thunman2007}
M.~Thunman, S.~Eckert, O.~Hennig, J.~Björkvall, and Du~Sichen.
\newblock Study on the formation of open-eye and slag entrainment in gas
  stirred ladle.
\newblock {\em steel research international}, 78(12):849--856, 2007.

\bibitem{Lv2017}
Ning ning Lv, Liu shun Wu, Hai chuan Wang, Yuan chi Dong, and Chang Su.
\newblock Size analysis of slag eye formed by gas blowing in ladle refining.
\newblock {\em Journal of Iron and Steel Research, International},
  24(3):243--250, 2017.

\bibitem{Valentin2009}
P.~Valentin et~al.
\newblock Influence of the stirring gas in a 170‐t ladle on mixing
  phenomena--formation and on‐line control of open‐eye at an industrial ld
  steel plant.
\newblock {\em steel research international}, 80(8):552--558, 2009.

\bibitem{Liu2011}
Heping Liu, Zhenya Qi, and Mianguang Xu.
\newblock Numerical simulation of fluid flow and interfacial behavior in
  three‐phase argon‐stirred ladles with one plug and dual plugs.
\newblock {\em Steel research international}, 82(4):440--458, 2011.

\bibitem{Li2008}
Baokuan Li et~al.
\newblock Modeling of three-phase flows and behavior of slag/steel interface in
  an argon gas stirred ladle.
\newblock {\em ISIJ international}, 48(12):1704--1711, 2008.

\bibitem{Li2015}
Linmin Li, Zhongqiu Liu, Baokuan Li, Hiroyuki Matsuura, and Fumitaka
  Tsukihashi.
\newblock Water model and cfd-pbm coupled model of gas-liquid-slag three-phase
  flow in ladle metallurgy.
\newblock {\em ISIJ International}, 55(7):1337--1346, 2015.

\bibitem{Li2016}
Linmin Li and Baokuan Li.
\newblock Investigation of bubble-slag layer behaviors with hybrid
  {Eulerian--Lagrangian} modeling and large eddy simulation.
\newblock {\em JOM}, 68(8):2160--2169, 2016.

\bibitem{Li2017}
Linmin Li, Baokuan Li, and Zhongqiu Liu.
\newblock Modeling of gas-steel-slag three-phase flow in ladle metallurgy: Part
  ii. multi-scale mathematical model.
\newblock {\em ISIJ International}, (ISIJINT-2017), 2017.

\bibitem{Singh2016}
Umesh Singh et~al.
\newblock Multiphase modeling of bottom-stirred ladle for prediction of
  slag–steel interface and estimation of desulfurization behavior.
\newblock {\em Metallurgical and Materials Transactions B}, 47(3):1804--1816,
  2016.

\bibitem{Ramasetti2019}
Eshwar~Kumar Ramasetti et~al.
\newblock Physical and cfd modeling of the effect of top layer properties on
  the formation of open‐eye in gas‐stirred ladles with single and
  dual‐plugs.
\newblock {\em steel research international}, 90(8):1900088, 2019.

\bibitem{Jonsson1996}
Lage Jonsson and Pär Jönsson.
\newblock Modeling of fluid flow conditions around the slag/metal interface in
  a gas-stirred ladle.
\newblock {\em ISIJ international}, 36(9):1127--1134, 1996.

\bibitem{Jonsson1998}
Lage Jonsson, Du~Sichen, and Pär Jönsson.
\newblock A new approach to model sulfurrefining in a gas-stirred ladle--a
  coupled cfd and thermodynamic model.
\newblock {\em ISIJ international}, 38(3):260--267, 1998.

\bibitem{Lejeune1995}
Anne‐Marie Lejeune and Pascal Richet.
\newblock Rheology of crystal‐bearing silicate melts: An experimental study
  at high viscosities.
\newblock {\em Journal of Geophysical Research: Solid Earth},
  100(B3):4215--4229, 1995.

\bibitem{Launder1972}
B.~E. Launder and D.~B. Spalding.
\newblock {\em Lectures in Mathematical Models of Turbulence}.
\newblock Academic Press, London, England, 1972.

\bibitem{Geng2010}
Dian-Qiao Geng, Hong Lei, and Ji-Cheng He.
\newblock Numerical simulation for collision and growth of inclusions in ladles
  stirred with different porous plug configurations.
\newblock {\em ISIJ International}, 50(11):1597--1605, 2010.

\bibitem{Llanos2010}
Carlos~A. Llanos, Saul Garcia, J.~Angel Ramos-Banderas, Jose de~J.~Barreto, and
  Gildardo Solorio.
\newblock Multiphase modeling of the fluidynamics of bottom argon bubbling
  during ladle operations.
\newblock {\em ISIJ International}, 50(3):396--402, 2010.

\bibitem{Zhang2012}
Lifeng Zhang.
\newblock Application of computational fluid dynamics (cfd) modeling on the
  transport phenomena during casting process.
\newblock {\em JOM}, 64(9):1059--1062, 2012.

\bibitem{Haojian2018}
Duan Haojian, Lifeng Zhang, Brian Thomas, and Alberto Conejo.
\newblock Fluid flow, dissolution, and mixing phenomena in argon-stirred steel
  ladles.
\newblock {\em Metallurgical and Materials Transactions B}, 49, 07 2018.

\bibitem{Batchelor1967}
G.~K. Batchelor.
\newblock {\em An Introduction to Fluid Dynamics}.
\newblock Cambridge University Press, Cambridge, England, 1967.

\bibitem{Roscoe1952}
R.~Roscoe.
\newblock The viscosity of suspensions of rigid spheres.
\newblock {\em British journal of applied physics}, 3(8):267, 1952.

\bibitem{Marsh1981}
B.~D. Marsh.
\newblock On the crystallinity, probability of occurrence, and rheology of lava
  and magma.
\newblock {\em Contributions to Mineralogy and Petrology}, 78(1):85--98, 1981.

\bibitem{bale2002factsage}
Christopher~W Bale, Patrice Chartrand, SA~Degterov, G~Eriksson, K~Hack, R~Ben
  Mahfoud, J~Melan{\c{c}}on, AD~Pelton, and S~Petersen.
\newblock Factsage thermochemical software and databases.
\newblock {\em Calphad}, 26(2):189--228, 2002.

\bibitem{Sinha2020}
A.~Sinha.
\newblock Thesis.
\newblock Master's thesis, 2020.

\bibitem{Rosenberg2005}
C.~L. Rosenberg and M.~R. Handy.
\newblock Experimental deformation of partially melted granite revisited:
  implications for the continental crust.
\newblock {\em Journal of Metamorphic Geology}, 23(1):19--28, 2005.

\bibitem{Glaser2012}
Björn Glaser.
\newblock {\em A Study on the Thermal State of Steelmaking Ladles}.
\newblock PhD thesis, KTH Royal Institute of Technology, 2012.

\bibitem{Wilcox1988}
David~C. Wilcox.
\newblock Reassessment of the scale-determining equation for advanced
  turbulence models.
\newblock {\em AIAA Journal}, 26(11):1299--1310, 1988.

\bibitem{Danckwerts1951}
P.~V. Danckwerts.
\newblock Significance of liquid-film coefficients in gas absorption.
\newblock {\em Ind. Eng. Chem.}, 43(6):1460--1467, 1951.

\bibitem{Issa1986}
Raad~I. Issa, A.~D. Gosman, and A.~P. Watkins.
\newblock The computation of compressible and incompressible recirculating
  flows by a non-iterative implicit scheme.
\newblock {\em Journal of Computational Physics}, 62(1):66--82, 1986.

\bibitem{Mavriplis2003}
Dimitri Mavriplis.
\newblock Revisiting the least-squares procedure for gradient reconstruction on
  unstructured meshes.
\newblock In {\em 16th AIAA Computational Fluid Dynamics Conference}, 2003.

\bibitem{Fluent2011}
ANSYS Inc., USA.
\newblock {\em Ansys Fluent Theory Guide}, 2011.

\bibitem{Lamont1970}
John~C. Lamont and D.~S. Scott.
\newblock An eddy cell model of mass transfer into the surface of a turbulent
  liquid.
\newblock {\em AIChE Journal}, 16(4):513--519, 1970.

\end{thebibliography}

\end{document}